\documentclass[a4paper,11pt]{article}
\usepackage{jheppub}
\usepackage[T1]{fontenc}
\usepackage{hyperref}
\usepackage{tensor}
\usepackage{epstopdf}
\usepackage{epsfig}
\usepackage{subfigure}
\usepackage[dvipsnames]{xcolor}
\usepackage[title]{appendix}
\usepackage{amsmath,epsfig}
\usepackage{amssymb,amsfonts}
\usepackage{latexsym}
\usepackage{bbold}
\usepackage{subeqnarray}
\usepackage{xcolor}
\usepackage{graphicx}
\usepackage{longtable}
\usepackage{color}
\usepackage{hyperref}
\usepackage{cleveref}
\usepackage{float}
\usepackage{pdfpages}
\usepackage{pdflscape}
\usepackage{caption}
\hypersetup{
      colorlinks=true,
      filecolor=black,
      anchorcolor=blue,
      linkcolor=blue,
      citecolor=blue,
      urlcolor=blue,
      linktocpage=true,
      plainpages=false,
      breaklinks=true,
      pdfstartview=FitH
           }

\relax
\DeclareMathOperator\arctanh{arctanh}

\usepackage{changes}
    \definechangesauthor[name={Per cusse}, color=orange]{per}

\begin{document}
\title{Towards classifying the interior dynamics of charged black holes with scalar hair}

\author{Rong-Gen Cai$^{1,2,3}$, Mei-Ning Duan$^{2,4}$, Li Li$^{1,2,4,5}$, Fu-Guo Yang$^{1,4}$}

\affiliation{$^{1}$School of Fundamental Physics and Mathematical Sciences, Hangzhou Institute for Advanced Study, UCAS, Hangzhou 310024, China.}

\affiliation{$^{2}$CAS Key Laboratory of Theoretical Physics, Institute of Theoretical Physics, Chinese Academy of Sciences, P.O. Box 2735, Beijing 100190, China.}

\affiliation{$3$School of Physical Science and Technology, Ningbo University, Ningbo, 315211, China}

\affiliation{$^{4}$School of Physical Sciences, University of Chinese Academy of Sciences, No.19A Yuquan Road, Beijing 100049, China.}

\affiliation{$^5$Peng Huanwu Collaborative Center for Research and Education, Beihang University, Beijing 100191, China.}

\emailAdd{cairg@itp.ac.cn}
\emailAdd{duanmeining@itp.ac.cn}
\emailAdd{liliphy@itp.ac.cn}
\emailAdd{yangfuguo@ucas.ac.cn}

\abstract{
The study of the interior of hairy black holes has received significant attention recently. This paper builds upon our recent analytical approach to investigate the internal dynamics of charged black holes with scalar hair in general spacetime dimensions. The geometries of these hairy balck holes end at a spacelike singularity. We investigate the alternation of Kasner epoch at later interior times and obtain the analytic expression for two kinds of transformation, namely Kasner inversion and Kasner transition. Moreover, we classify three different types of Kasner alternations for a large class of Einstein-Maxwell-scalar theory. Our analytical results are corroborated by numerical solutions to the full equations of motion, including a top-down model from supergravity. For general interactions, more complicated behaviors beyond our analytical description are also found and discussed, including the presence of non-Kasner epochs and the random change of the amplitude of the Kasner exponent at late interior times.}

\maketitle
\flushbottom

\noindent
\newpage

\section{Introduction}\label{Intro}

Identifying the interior structures and their underlying dynamics is an important step toward understanding the nature of black holes. In particular, the appearance of the inner Cauchy horizon of a black hole results in breaking down classical predictability and appears to violate strong cosmic censorship. Despite decades of extensive efforts, even in general relativity, a complete understanding of the interior of black holes remains elusive. Rich classical dynamics have been uncovered over the past few decades, in particular, the emergence of Belinskii-Khalatnikov-Lifshitz (BKL) chaos~\cite{Belinsky:1970ew,Belinski:1973zz}. Based on the BKL hypothesis, the dynamics in the vicinity of a spacelike singularity can be asymptotically described as billiard motion in a region of Lobachevskii space, which is known as "Cosmological Billiards"~\cite{Damour:2002et}. However, our theoretical understanding of black hole interior from Cosmological Billiards is also incomplete, and it is even more so when considering matter content with general interactions.

Stimulated by the holographic duality, there has been growing interest towards exploring the internal structure of a black hole in recent years. The authors of~\cite{Frenkel:2020ysx,Hartnoll:2020rwq} considered a free scalar in the neutral AdS black hole, which corresponds to turning on a relevant scalar operator of the dual thermal CFT state. They found that there is in general no Cauchy horizon and, at late interior times, the spacetime dives into a stable Kasner geometry. The case with a free charged scalar field was considered in~\cite{Hartnoll:2020fhc}, known as the holographic superconductor.
Some intricate behaviors were found before ending at a spacelike Kasner singularity, including the collapse of the Einstein-Rosen (ER) bridge, the Josephson oscillations of the scalar field and possible alternation between neighbouring Kasner epochs. More rigorous ``no Cauchy horizon theorem'' of a black hole with (charged) scalar hair was proven in~\cite{Cai:2020wrp,An:2021plu} by constructing a radially conserved "charge" and the null energy condition (see~\cite{Devecioglu:2021xug,Devecioglu:2023hmn} for generalization). Interestingly, without referring to the form of matter fields, the number of horizons of static black holes can be strongly constrained by energy conditions~\cite{Yang:2021civ}.

The interior dynamics of scalarized black holes has subsequently been studied in the literature, including the variation of interactions~\cite{Grandi:2021ajl,Liu:2021hap,Dias:2021afz,An:2022lvo,Hartnoll:2022rdv}, additional matter content~\cite{Mansoori:2021wxf,Sword:2021pfm,Mirjalali:2022wrg}, and analysis of RG flows \cite{Wang:2020nkd,Caceres:2022smh,Caceres:2022hei}. The generalization to anisotropic case can be found in the cases with vector hair~\cite{Cai:2021obq,Sword:2022oyg}, helical structure~\cite{Liu:2022rsy}, as well as holographic topological semimetals~\cite{Gao:2023zbd}. Nevertheless, what is the duality of the interior dynamics in field theory remains a fascinating and challenging problem if the holographic principle is considered as the basic principle in physics. 

Because the nonlinear effect plays an important role inside the black holes, the internal dynamics obviously depends on the details of the model one considers. So far, interesting internal behaviors have been observed numerically in some specific models.
In particular, at late interior times, a common feature is the emergence of Kasner scaling towards the singularity. Depending on the interactions, a further phenomenon appearing in these works is bounces between different Kasner epochs. In this paper we aim at classifying the interior of charged black holes with non-trivial scalar hair, with the key aspects of the dynamics captured analytically. Moreover, previous studies mainly focus on the hairy black holes in four spacetime dimensions. We would like to consider general dimensions since there could be some interesting dynamics that appear in higher dimensions.\,\footnote{For example, the Reissner-Nordstr\"{o}m de Sitter black holes are linear unstable to gravitational perturbations only in six spacetime dimensions and above~\cite{Dias:2020ncd} and can become unstable in five dimensions and above with Gauss-Bonnet correction~\cite{Cai:2021qcq}.} 

More precisely, we will consider the generalized Einstein-Maxwell-scalar theory that allows scalar coupling with less restrictions and provides a general scalar theory with local $U(1)$ symmetry. For scalar couplings with a polynomial form, we can find that there will eventually be a Kasner singularity. Before ending the singularity, the interior can have the "Kasner inversion" behavior in some cases and the alternation law between the two Kanser epochs is given analytically. When the exponential coupling term is introduced, it will lead to the transformation behavior called "Kasner transition" which can also be described analytically. Three classes of alternation of Kasner epochs will be provided and will be numerically verified. In addition, we will explore more complicated cases with general couplings and scalar potentials, for which some novel internal dynamics beyond all known analytical description will be shown.

The paper is organized as follows. In Section~\ref{Sec:model}, we introduce the gravitational model and establish the equations of motion. In Section~\ref{Sec:GeneralAnalysis}, we discuss the alternation of Kasner epochs at late interior times in general spacetime dimensions, for which, under certain approximations, we are able to obtain self-consistent asymptotic solutions. Numerical verification is provided by considering some benchmark examples. More cases with general couplings and potentials are discussed in Section~\ref{Sec:general}. We conclude with some discussions in Section~\ref{Sec:ConAndDis}.

\section{Setup}\label{Sec:model}
We start with a (d+1)-dimensional theory that admits general scalar couplings and 
local $U(1)$ symmetry:
\begin{equation}\label{Model}
\begin{split}
S=&\frac{1}{2\kappa_N^2}\int d^{d+1}x\sqrt{-g} \left[R+\mathcal{L}\right]\,,\\
\mathcal{L}&=-\frac{1}{2}(\partial_\mu\psi)^2-\mathcal{F}(\psi)(\partial_\mu\theta-q A_\mu)^2-V(\psi)-\frac{Z(\psi)}{4}F_{\mu\nu}F^{\mu\nu}
\end{split}
\end{equation}
where $\psi$ and $\theta$ are both real scalars, and $A_\mu$ is the $U(1)$ gauge field with its strength $F_{\mu\nu}=2\partial_{[\mu} A_{\nu]}$. The three couplings $\mathcal{F}$, $V$ and $Z$ depend on $\psi$ and can take quite general form. We only require $\mathcal{F}$ and $Z$ to be positive to ensure positivity of the kinetic term for $\theta$ and $A_\mu$. Depending on the scalar potential $V$ (as well as $Z$ and $\mathcal{F}$), the spacetime can be asymptotically flat, anti-de Sitter (AdS), dS or other geometries. In AdS case, such generalized St\"{u}ckelberg theory has been used to realize the superconducting phase transition in holography (see \emph{e.g.}~\cite{Franco:2009yz,Kiritsis:2015hoa}).

We wish to study the hairy black hole solutions that take the form
\begin{equation}\label{MetricFieldAnsatz}
\mathrm{d}s^2=\frac{1}{z^2}\left[-f(z)\mathrm{e}^{-\chi(z)}\mathrm{d}t^2+\frac{\mathrm{d}z^2}{f(z)}+\mathrm{d}\Sigma^2_{d-1,k}\right],\quad \psi=\psi(z),\quad A=A_t(z) \mathrm{d}t\,,
\end{equation}
with $z$ the radial coordinate. Here $\mathrm{d}\Sigma^2_{d-1,k}$ denotes the metric of unit sphere $(k=1)$, planar $(k=0)$ or unit hyperbolic plane $(k=-1)$ in $(d-1)$-dimensions. Moreover, we have chosen $\theta=0$ without loss of generality. Then, the equations of motion are given as follows.
\begin{equation}\label{EoM:psi}
\psi''=-\left(\frac{1}{z}+\frac{h'}{h}\right)\psi'-\frac{q^2A_t^2}{z^{2d}h^2}\frac{\mathrm{d}\mathcal{F}}{\mathrm{d}\psi}+\frac{\mathrm{e}^{-\chi/2}}{z^{d+2}h}\frac{\mathrm{d}V}{\mathrm{d}\psi}+\frac{\mathrm{e}^{\chi/2}}{2z^{d-2}h}\frac{\mathrm{d}Z}{\mathrm{d}\psi}\,,
\end{equation}
\begin{equation}\label{EoM:at}
\left(\frac{\mathrm{e}^{\chi/2}ZA_t'}{z^{d-3}}\right)'=\frac{2q^2A_t}{z^{2d-1}h}\mathcal{F}\,,
\end{equation}
\begin{equation}\label{EoM:chi}
(d-1)\chi'=z\psi'^2+\frac{2q^2A_t^2}{z^{2d-1}h^2}\mathcal{F}\,,
\end{equation}
\begin{equation}\label{EoM:h}
h'=\frac{\mathrm{e}^{-\chi/2}}{d-1}\left(-\frac{k(d-1)(d-2)}{z^{d-1}}+\frac{V}{z^{d+1}}+\frac{\mathrm{e}^{\chi}}{2z^{d-3}}ZA_t'^2\right),
\end{equation}
where the prime denotes the derivative with respect to $z$ and we have introduced $h=z^{-d}\mathrm{e}^{-\chi/2}f$ for later convenience. 

In our coordinate system, the boundary is at $z=0$ and the singularity would be at $z\rightarrow\infty$. Denoting the event horizon as $z_H$ at which $f(z)$ vanishes, the temperature and entropy density can be obtained as
\begin{equation}
 T=-\frac{\mathrm{e}^{-\chi\left(z_H\right) / 2} f^{\prime}\left(z_H\right)}{4 \pi},\quad s=\frac{2\pi}{\kappa_N^2 z_H^{d-1}}\,.  
\end{equation}
All the functions of~\eqref{MetricFieldAnsatz} are continuous near the horizon. In particular, one has $A_t(z_H)=0$ once $q\neq 0$ of~\eqref{Model}. The boundary condition away from the event horizon depends on the asymptotics of spacetime. A large number of hairy black hole solutions to the above equations have been constructed numerically in the case of asymptotic AdS and flatness. In contrast, a no scalar hair theorem for charged black holes in dS spacetime has been recently proved~\cite{An:2023bpb}.  

While the solutions outside the event horizon depend on the details of the couplings, it has been shown that the hairy black holes~\eqref{MetricFieldAnsatz} to the theory~\eqref{Model} do not have an inner Cauchy horizon~\cite{An:2022lvo} proved by using a radially conserved charge~\cite{Cai:2020wrp} and the null energy condition~\cite{An:2021plu}. The internal dynamics would end at a spacelike singularity. In the following, we shall classify the interior dynamics of those hairy black holes. The collapse of the ER bridge and the scalar oscillations are closely related to the instability of the would-be inner Cauchy horizon triggered by the scalar hair. Both appear near the would-be inner Cauchy horizon and are sensitive to the temperature. They become less dramatic and finally disappear as the temperature is kept away from the critical temperature $T_c$ at which the scalar hair develops. On the other hand, the presence of Kasner epoch in deep interior is a more robust feature and the possible alternation of different Kasner epochs deserves a better understanding. Recently, we have obtained analytically the transformation rule for the alternation of Kasner epochs in a top-down holographic superconductor~\cite{An:2022lvo}, which provides some useful tools for further research. Therefore, in this work we focus on the alternation law of different Kasner epochs at late time evolution of the interior. Moreover, we will uncover the internal dynamics beyond the Kasner scaling.

\section{Kasner Structure and Alternation}
\label{Sec:GeneralAnalysis}
Due to the strong nonlinear nature of the equations of motion~$\eqref{EoM:psi}$-$\eqref{EoM:h}$, it is impossible to solve the system analytically. Nevertheless, under certain approximations, we are able to obtain self-consistent asymptotic solutions. This procedure will be further established by checking the full numerical solutions. To simplify our analysis, we shall set $Z=1$ in the present study. Our strategy is to take a Kasner regime as background and to consider possible deviation that may lead to the alternation to another Kasner epoch.

\subsection{Kasner Epoch}\label{subKasner}
Le's start with the simple case in which the contributions to $\eqref{EoM:psi}$-$\eqref{EoM:h}$ from 
$\mathcal{F}$ and $V$ are negligible. For example, both are polynomial functions. Then, for $d\geq3$, the approximate differential equations at large $z$ in deep interior could be
\begin{equation}\label{ApproximateEoM}
\begin{aligned}
\psi''=-\frac{1}{z}&\psi'\,,\quad \left(\frac{\mathrm{e}^{\chi/2}A_t'}{z^{d-3}}\right)'=0\,,\quad (d-1)\chi'=z\psi'^2\,,\\
&h'=\frac{1}{2(d-1)}\left(\frac{\mathrm{e}^{\chi/2}A_t'}{z^{d-3}}\right)^2 z^{d-3}\mathrm{e}^{-\chi/2}\,,
\end{aligned}
\end{equation}
where we have dropped all the terms associated with $\mathcal{F}$ and $V$. 

With these approximations, one can explicitly solve~\eqref{ApproximateEoM} and obtain
\begin{equation}
\begin{split}\label{ApproximateSolutions}
\psi=\alpha \ln z +C_{\psi}\,,\qquad \chi=\frac{\alpha^2}{d-1}\ln z +C_{\chi}\,,\\ A_t'=C_{A_t}z^{d-3}\mathrm{e}^{-\chi/2}\,,\qquad h'=C_h z^{d-3}\mathrm{e}^{-\chi/2}\,,
\end{split}
\end{equation}
where $\alpha$, $C_{\psi}$, $C_{A_{t}}$, $C_{\chi}$ and $C_{h}$ are integral constants. In particular, $C_h>0$ from the last equation of~\eqref{ApproximateEoM}. Meanwhile, we have also assumed that $h'$ is integrable, \emph{i.e.} the order of $h'/h$ is smaller than $1/z$ so that it could be neglected in~\eqref{EoM:psi}. When the integrability assumption about $h'$ is broken, a new dynamic process called Kasner inversion will develop, which we will discuss below. Before that, let's understand the asymptotic solutions~$\eqref{ApproximateSolutions}$ first. 

From the solution~$\eqref{ApproximateSolutions}$, the background in the deep interior is given by
\begin{equation}
\mathrm{d}s^2=\frac{1}{z^2}\left[-\frac{\mathrm{d}z^2}{C_f z^d\mathrm{e}^{\chi/2}}+C_f z^d\mathrm{e}^{-\chi/2}\mathrm{d}t^2+\mathrm{d}\Sigma^2_{d-1,k}\right],\quad \psi\simeq\alpha\ln z\,,
\end{equation}
in which all the metric components are power laws and the scalar field is logarithmic. 
After converting to the proper time $\tau\sim z^{-(\frac{d}{2}+\frac{\alpha^2}{4(d-1)})}$, one obtains
\begin{equation}\label{myKasnerU}
\mathrm{d}s^2=-\mathrm{d}\tau^2+c_t \tau^{2p_t}\mathrm{d}t^2+c_s\tau^{2p_s}\mathrm{d}\Sigma^2_{d-1,k},\quad \psi\simeq -\sqrt{2}p_{\psi}\ln \tau\,,
\end{equation}
where
\begin{equation}
p_t=\frac{\alpha^2-2(d-1)(d-2)}{\alpha^2+d(d-1)},\quad p_s=\frac{4(d-1)}{\alpha^2+d(d-1)},\quad p_\psi=\frac{2\sqrt{2}(d-1)\alpha}{\alpha^2+d(d-1)}\,,
\end{equation}
with $c_{t}$ and $c_{s}$ constants. One immediately finds that
\begin{equation}
  p_t+(d-1)p_s=1,\quad p_t^2+(d-1)p_s^2+p_{\psi}^2=1\,,
\end{equation}
and thus the geometry is equipped with the Kasner structure. Notice that $\alpha=z\psi'$ is a constant and determines other exponents in a Kasner universe~\eqref{myKasnerU}. We shall call $\alpha$ the Kasner exponent. The Schwarzschild singularity is obtained by taking $\alpha=0$.

A natural question is whether our approximate solution~$\eqref{ApproximateSolutions}$ makes sense or not. Therefore, we should check if the terms we discarded are small in a given Kasner universe. In the equations of motion~$\eqref{EoM:psi}$-$\eqref{EoM:h}$, with approximate solution~$\eqref{ApproximateSolutions}$, one obtains the following constraints:
\begin{equation}\label{constraint}
\begin{aligned}
\mathcal{O}\left(\frac{q^2A_t^2}{z^{2d}h^2}\frac{\mathrm{d}\mathcal{F}}{\mathrm{d}\psi}\right)&<\mathcal{O}\left(\frac{1}{z^{2d-1}}\right),\quad
\mathcal{O}\left(\frac{\mathrm{e}^{-\chi/2}}{z^{d+2}h}\frac{\mathrm{d}V}{\mathrm{d}\psi}\right)<\mathcal{O}\left(\frac{1}{z^{d+1}}\right),\\
\mathcal{O}\left(\frac{q^2A_t}{z^{2d-1}h}\mathcal{F}\right)&<\mathcal{O}\left(\frac{1}{z^{2d-2}}\right),\quad \mathcal{O}\left(\frac{2q^2A_t^2}{z^{2d-1}h^2}\mathcal{F}\right)<\mathcal{O}\left(\frac{1}{z^{2d-2}}\right),\\
\mathcal{O}\left(\frac{\mathrm{e}^{-\chi/2}}{z^{d-1}}\right)&<\mathcal{O}\left(\frac{1}{z^{d-1}}\right),\quad \mathcal{O}\left(\frac{V\mathrm{e}^{-\chi/2}}{z^{d+1}}\right)<\mathcal{O}\left(\frac{1}{z^d}\right)\,,
\end{aligned}
\end{equation}
which allows $V$ and $\mathcal{F}$ to be arbitrary algebraic functions, including polynomial functions, as long as $d\geq3$.~\footnote{It is easy to check that, in the right hand of~\eqref{EoM:h}, the curvature term  due to the topology of the horizon is negligible compared to other terms when $d\geq3$.} Therefore, under~\eqref{constraint}, the neglected terms will not change the dynamic behaviors from the approximate equations~$\eqref{ApproximateEoM}$ and the approximate solution~$\eqref{ApproximateSolutions}$ is self-consistent. So far, $h'$ is still assumed to be integrable. 

Once above assumption is invalid, the solution~$\eqref{ApproximateSolutions}$ will become unstable towards the deep interior. A particularly simple case is triggered by the $h'/h$ term, resulting in the dynamics away from the unstable Kasner epoch. Interestingly, this alternation caused by the non-integrability of $h'$ will make itself come back to be integrable, and will enter the stable Kasner epoch finally. This process is called Kasner inversion.

\subsection{Kasner Inversion}\label{subsec:inversion}
The no-inner horizon theorem requires $h<0$ inside the event horizon. From~$\eqref{ApproximateSolutions}$, one finds that
\begin{equation}
 h'(z)\sim z^{d-3-\frac{\alpha^2}{2(d-1)}}\,.
\end{equation}
Therefore, to have a stable Kasner epoch all the way down to the singularity, the integral of $h'(z)$ should be finite, \emph{i.e.} $h'(z)$ is integrable. Otherwise, new dynamics will come into play, triggering the transformation to another epoch.

The breakdown of the integrability of $h'$ yields
\begin{equation}\label{KasInvCon}
 d-3-\frac{\alpha^2}{2(d-1)}>-1 \Rightarrow |\alpha|<\sqrt{2(d-1)(d-2)}\,.
\end{equation}
Under condition~$\eqref{KasInvCon}$, the background~$\eqref{ApproximateSolutions}$ will become unstable towards the singularity. In this case, one cannot drop the second term in parentheses in the scalar equation~\eqref{EoM:psi}, and the dynamics is controlled by the following equations:
\begin{equation}\label{KasInverEoM}
\frac{1}{z}(z\psi')'=-\frac{h'}{h}\psi', \quad h'=\frac{1}{2(d-1)}\left(\frac{\mathrm{e}^{\chi/2}A_t'}{z^{d-3}}\right)^2 z^{d-3}\mathrm{e}^{-\chi/2}\,.
\end{equation}
Notice that $h'$ is determined by the kinetic term of the gauge field.

Motivated by our previous work~\cite{An:2022lvo}, the above equations can be solved using the constant variant method. Let's assume that $\psi$ takes the form
\begin{equation}\label{psiPresolve}
\psi(z)=\int^z \frac{\widetilde\alpha(s)}{s}\mathrm{d}s\,.
\end{equation}
Substituting~$\eqref{psiPresolve}$ into the differential equations~\eqref{KasInverEoM}, one can obtain the following equation for $\widetilde\alpha(z)$:\,\footnote{More precisely, to obtain~\eqref{InvAlphaDiff} we have used the condition that the combination $\left(\frac{\mathrm{e}^{\chi/2}A_t'}{z^{d-3}}\right)$ of~\eqref{KasInverEoM} remains approximately a constant value over the range of the Kasner alternation. This condition is seen to hold at large $\psi$ for the kind of couplings we are considering.}
\begin{equation}\label{InvAlphaDiff}
  2z(d-1)\widetilde\alpha\widetilde\alpha''-4z(d-1)\widetilde\alpha'^2+\widetilde\alpha'\widetilde\alpha(\widetilde\alpha^2-2(d-1)(d-3))=0\,.
\end{equation}
Solving~\eqref{InvAlphaDiff}, one can obtain analytically that
\begin{equation}\label{InvAlphaSol}
\begin{split}
2(d-2)\ln\left[\frac{z}{z_{I}}\right]+\frac{2c_1 \sqrt{d-1} }{\sqrt{(d-1)c_1^2-2d+4}}\arctanh\left[\frac{c_1(d-1) -\widetilde\alpha[z]}{\sqrt{d-1} \sqrt{(d-1)c_1^2-2d+4}}\right]\\
+2\ln\left|\frac{\widetilde\alpha[z]}{c_1(d-1)}\right|+2\ln\left|\frac{c_1^2(d-1)^2-2(d-1)(d-2)}{\widetilde\alpha[z]^2-2(d-1) c_1 \widetilde\alpha[z]+2(d-1)(d-2)}\right|=0\,,
\end{split}
\end{equation}
where $c_1$ and $z_I$ are constants with $z_I$ satisfying $\widetilde\alpha[z_I]=c_1(d-1)$.
The value of $\alpha$ for the Kasner epoch before (after) the transformation is obtained by taking the limit $z/z_I\ll 1$ ($z/z_I\gg 1$).

As an implicit function, it is not easy from~\eqref{InvAlphaSol} to obtain the relation of the Kasner exponents for the Kasner inversion. A convenient method is as follows. One first observes that both the arctanh term and the last term of~\eqref{InvAlphaSol} go to infinity at the same time, since
\begin{equation}
   \left|\frac{c_1(d-1) -\widetilde\alpha[z]}{\sqrt{d-1} \sqrt{(d-1)c_1^2-2d+4}}\right|\rightarrow 1 \Leftrightarrow \widetilde\alpha[z]^2-2(d-1) c_1 \widetilde\alpha[z]+2(d-1)(d-2)\rightarrow 0\,.
\end{equation}
As a consequence, when $\ln[z/z_{I}]$ goes to infinity, to make the equation~\eqref{InvAlphaSol} valid, both the arctanh term and the last term of~\eqref{InvAlphaSol} should go to infinity as offset.~\footnote{Note that $\widetilde\alpha$ should not divergent to avoid any singularity at which the spacetime terminates. Thus, the third term of~\eqref{InvAlphaSol} is finite.} 

As $z$ goes from $z/z_{I}\ll 1$ to $z/z_{I}\gg 1$, $\ln[z/z_{I}]$ changes from $-\infty$ to $+\infty$, which means that the arctanh term and the last term of~\eqref{InvAlphaSol} change from $+\infty$ to $-\infty$. Therefore, the two exponents $\alpha$ for the Kasner epochs before and after the Kasner inversion are the roots of the following quadratic equation for $\widetilde\alpha$:
\begin{equation}
    \widetilde\alpha^2-2 c_1(d-1) \widetilde\alpha+2(d-1)(d-2)=0\,.
\end{equation}
According to Vieta's Formula for quadratic equation, one immediately obtains the transformation law between two adjacent Kasner epochs.
\begin{equation}\label{KasInvLaw}
  \alpha\alpha_{I}=2(d-1)(d-2)\,,
\end{equation}
where $\alpha$ is the Kasner exponent before the Kasner inversion, and $\alpha_{I}$ is the one after the inversion.

Suppose that in a certain Kasner epoch $|\alpha|<\sqrt{2(d-1)(d-2)}$ for which $h'$ is not integrable, see~\eqref{KasInvCon}. One immediately finds from~\eqref{KasInvLaw} that $|\alpha_I|=2(d-1)(d-2)/|\alpha|>\sqrt{2(d-1)(d-2)}$. Therefore, $h'$ becomes integrable after the Kasner inversion process and gives a stable Kasner epoch. Thus, the Kasner inversion provides a stable mechanism for Kasner dynamics. In addition, one can find that the transformation law of Kasner inversion~$\eqref{KasInvLaw}$ depends only on the spacetime dimension, which is different from the Kasner transition we discuss in the next Subsection.

In addition, to understand this transformation qualitatively, one can consider the dominant  term in the Kasner inversion process, for which the equation of motion with respect to $\psi$ reads
\begin{equation}\label{hDominates}
\psi''=-\left(\frac{1}{z}+\frac{h'}{h}\right)\psi'\,.
\end{equation}
Note that inside the event horizon $h<0$  and $h'>0$ from the last equation of~\eqref{KasInverEoM}. Then, substituting~\eqref{psiPresolve} into the above equation, one has
\begin{equation}
\widetilde\alpha'\widetilde\alpha=-\frac{h'}{h}\widetilde\alpha^2>0\,.
\end{equation}
In other words, once $\widetilde\alpha$ is positive in a Kasner epoch, the kinetic term of the gauge field will cause $\widetilde\alpha'>0$, leading to an increase in $\widetilde\alpha$ as $z$ is increased. On the other hand, as we have shown in this Subsection, the increase in $\widetilde\alpha$ will further cause $h'$ to be away from the non-integrability condition~\eqref{KasInvCon}. The same discussion applies to the case with negative $\widetilde\alpha$. Therefore, although the kinetic energy of the gauge field causes the instability of a Kasner epoch when its exponent $\alpha$ is within~\eqref{KasInvCon}, it would not completely destroy the Kasner structure, but causes the unstable Kasner epoch to transform into a Kasner epoch with a larger value of $\alpha$ given by~\eqref{KasInvLaw}. In the new Kasner epoch, $h'$ is integrable, thus is stable if no other dynamics come into play. Interestingly, we will see similar stable mechanisms in the Kasner transition and even the Kasner transformation caused by the scalar potential.

\subsection{Kasner Transiton}\label{SubSec:KasTrans}
Many dimensional reductions of superstring/supergravity theory lead to exponential couplings for the various Kaluza-Klein scalar fields. We therefore consider the coupling function $\mathcal{F}$ that takes an exponential form $\mathcal{F}(\psi)\sim \mathrm{e}^{\kappa\psi}$ asymptotically. 
We will show that the self-consistency of the solution~$\eqref{ApproximateSolutions}$ might be destroyed, and a new Kasner transformation process will appear. 

We begin with the Kasner solution~$\eqref{ApproximateSolutions}$ and assume that $h'$ is integrable. Therefore, when the order of the coupling term satisfies
\begin{equation}\label{KasTranCon}
\mathcal{O}\left(\frac{A_t^2}{z^{2d}h^2}\frac{\mathrm{d}\mathcal{F}}{\mathrm{d}\psi}\right)=\mathcal{O}\left(\frac{z^{\kappa\alpha}}{z^{2d}}\right)\geq\mathcal{O}\left(\frac{\psi'}{z}\right)\,,
\end{equation}
or equivalently
\begin{equation}\label{LowTransLimit}
\kappa\alpha>(2d-2)\,,
\end{equation}
one can not drop the second term $\sim\frac{\mathrm{d}\mathcal{F}}{\mathrm{d}\psi}$ in the right hand of~\eqref{EoM:psi}. Then,
the equation~\eqref{EoM:psi} is approximated by~\footnote{In contrast, when $\kappa\alpha<(2d-2)$, the right hand side of~\eqref{KasTranEoM} is approximatively zero. Therefore, we do not expect to have an alternation to a new Kasner epoch.}
\begin{equation}\label{KasTranEoM}
  \frac{1}{z}(z\psi')'\simeq \frac{\mathrm{e}^{\kappa \psi}}{z^{2d}}\,.
\end{equation}
Note that we have used the fact that $A_t$ and $h$ are at the same order, as can be seen from~\eqref{ApproximateSolutions}. We will furthermore verify that the expressions we obtain under the above assumption agree with numerical results.

Given that~\eqref{ApproximateEoM} is still a good approximation, the equation~\eqref{KasTranEoM} can be solved in terms of the constant variant method as in~\eqref{psiPresolve}. 
We then find the following differential equation for $\widetilde{\alpha}$.
\begin{equation}\label{TranAlphaDiff}
z \widetilde{\alpha}''+(2d-1)\widetilde\alpha'-\kappa\widetilde\alpha\widetilde\alpha'=0\,.
\end{equation}
Solving the equation~\eqref{TranAlphaDiff} yields
\begin{equation}
\widetilde{\alpha}(z)=\frac{2 d-2-c_1 \operatorname{tanh}\left[c_1\ln(z/z_T)\right]}{\kappa}\,,
\end{equation}
where $c_1$ and $z_T$ are integration constants. The latter denotes the position in the transition region with $\widetilde{\alpha}(z_T)=(2d-2)/\kappa$.
The value of $\alpha$ in the Kasner epoch before the transition is obtained by taking $z/z_T\ll 1$, \emph{i.e.} $\alpha=(2d-2+c_1)/\kappa$. The one after the transition is determined by taking $z/z_T\gg 1$, \emph{i.e.} $\alpha_T=(2d-2-c_1)/\kappa$.
We then obtain the transformation law for the Kasner transition between two adjacent
Kasner epochs that is
\begin{equation}\label{KasTranLaw}
  \alpha +\alpha_T=\frac{2}{\kappa}(2d-2)\,.
\end{equation}

Supposing $\kappa \alpha>(2d-2)$ in a Kasner epoch, the law~\eqref{KasTranLaw} shows that the Kasner transition process will decrease the amplitude of $\alpha$ until the condition~\eqref{LowTransLimit} is destroyed. Moreover, the transformation law~$\eqref{KasTranLaw}$ of Kasner transition not only depends on the spacetime dimension, but also depends on the value of coupling constant $\kappa$ of $\mathcal{F(\psi)}$.

\subsection{Classification of Kasner Alternation}
\begin{figure}[H]
    \centering
    \begin{minipage}{0.32\textwidth}
        \centering
        \includegraphics[width=1\textwidth]{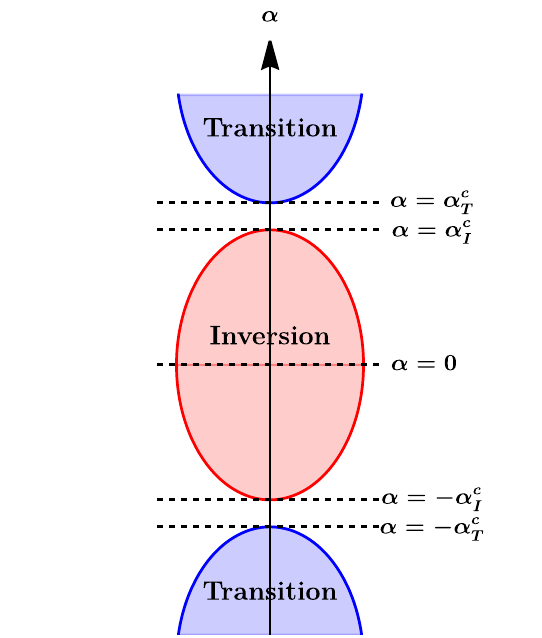}
        \end{minipage}
    \begin{minipage}{0.34\textwidth}
        \centering
        \includegraphics[width=1\textwidth]{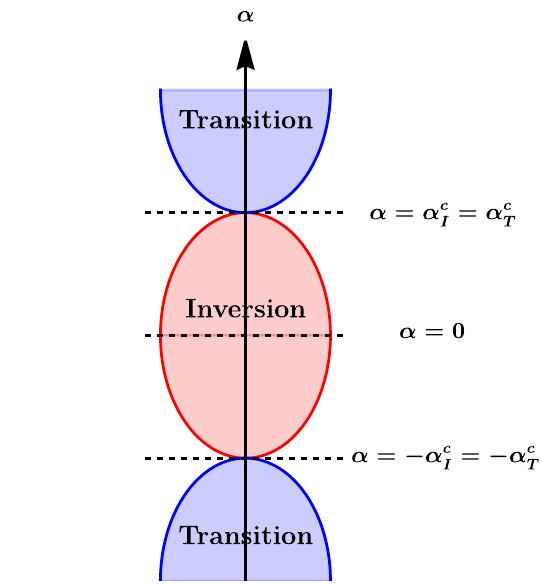}
    \end{minipage}
    \begin{minipage}{0.32\textwidth}
        \centering
        \includegraphics[width=1\textwidth]{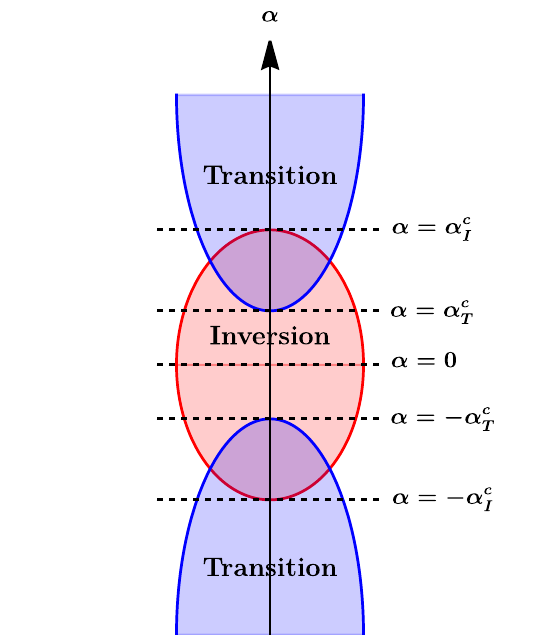}
    \end{minipage}
    \caption{Classification for the possible alternation of Kasner epochs for a theory with an exponential coupling $\mathcal{F}(\psi)\sim \mathrm{e}^{\kappa\psi}$. Give a Kasner eopch with the exponent $\alpha=z\psi'$ and denote $\alpha_I^c=\sqrt{2(d-1)(d-2)}$ and $\alpha_T^c=2(d-1)/|\kappa|$. A Kasner transition triggers when $\alpha$ falls into the blue region ($|\alpha|>\alpha_T^c$), and a Kanser inversion appears when $\alpha$ is in the red region ($|\alpha|<\alpha_I^c$). 
    The Kasner alternation can be divided into three classes depending on the spatial dimension $d$ and the coupling constant $\kappa$. \textbf{Left panel}: $\alpha_I^c<\alpha_T^c$. There exist a stable region with $\alpha_I^c<|\alpha|<\alpha_T^c$. 
    \textbf{Middle panel}: $\alpha_I^c=\alpha_T^c=\sqrt{2(d-1)(d-2)}$.  There will be an infinite sequence of Kasner alternations towards the singularity, except for the fine-tuning with $\alpha=\sqrt{2(d-1)(d-2)}$. \textbf{Right panel}: $\alpha_I^c>\alpha_T^c$. In the overlap of red and blue regions ($\alpha_T^c<|\alpha|<\alpha_I^c$), either Kasner transition or inversion description breaks down.}
    \label{fig:KasnerTI}
\end{figure}

From the above discussion, it can be seen that the Kasner transition causes the parameter $|\alpha|$ to decrease, while the Kasner inversion makes $|\alpha|$ to increase. When these two processes are triggered alternately, it could lead to an infinite chaotic oscillation of Kasner epochs. Therefore, for a theory with an exponential coupling $\mathcal{F}(\psi)\sim \mathrm{e}^{\kappa\psi}$, we have three different types of Kasner alternations
that are summarized in Fig.~\ref{fig:KasnerTI}. The red part in each panel of Fig.~\ref{fig:KasnerTI} denotes the case for Kasner inversion, while the blue one for the Kasner transition.

\paragraph{Case I:}
$\sqrt{2(d-1)(d-2)}<2(d-1)/|\kappa|$ (left panel of Fig.~\ref{fig:KasnerTI}). 

The Kasner transition occurs when $|\alpha|>2(d-1)/|\kappa|$ and the Kasner inversion occurs when $|\alpha|<\sqrt{2(d-1)(d-2)}$.  Once  $\sqrt{2(d-1)(d-2)}<|\alpha|<2(d-1)/|\kappa|$, both the Kasner transformations will not be triggered, thus the system settles down to a stable Kasner epoch. 

\paragraph{Case II:}

$\sqrt{2(d-1)(d-2)}=2(d-1)/|\kappa|$ (middle panel of Fig.~\ref{fig:KasnerTI}).

In this critical case, $|\alpha|=\sqrt{2(d-1)(d-2)}=2(d-1)/|\kappa|$ is the only fixed point. Therefore, for the initial value of $\alpha\neq \sqrt{2(d-1)(d-2)}=2(d-1)/|\kappa|$, there will be an infinite Kasner alternations towards the singularity. 

The benchmark model is the top-down theory in four dimensions we recently considered in~\cite{An:2022lvo}. Its Lagrangian reads
\begin{equation}\label{modelst4D}
\mathcal{L}^{(4)}=-\frac{1}{2}(\partial_\mu\psi)^2-\frac{\sinh^2\psi}{2}\left(\partial_\mu\theta-\frac{1}{L}A_\mu\right)^2+\frac{1}{L^2}\cosh^2\frac{\psi}{2}(7-\cosh\psi)-\frac{1}{4}F_{\mu\nu}F^{\mu\nu}\,,
\end{equation}
which is obtained as a consistent truncation of M-theory with $F\wedge F=0$. One has $\mathcal{F}=\frac{\sinh^2\psi}{2}\sim e^{2\psi}$ and therefore $\kappa=2, d=3$. One can check that our transformation laws~\eqref{KasInvLaw} and~\eqref{KasTranLaw} reduce precisely to the case in~\cite{An:2022lvo}. Indeed, a never-ending chaotic alternation of Kasner epochs towards the singularity was observed for the theory~\eqref{modelst4D} (see~\cite{An:2022lvo} for more details).

\paragraph{Case III:}

$\sqrt{2(d-1)(d-2)}>2(d-1)/|\kappa|$ (right panel of Fig.~\ref{fig:KasnerTI}).

When $|\alpha| >\sqrt{2(d-1)(d-2)}$, the Kasner transition develops, and when $|\alpha|<2(d-1)/\kappa$, the Kasner inversion appears. 

Nevertheless, for $2(d-1)/|\kappa|<|\alpha|<\sqrt{2(d-1)(d-2)}$ (the overlapping region in the right panel of Fig.~\ref{fig:KasnerTI}), both the contributions from $h'/h$ and $\mathcal{F}$ to~\eqref{EoM:psi} play important roles. The complex competition between the Kasner inversion and the transition could occur.
So far, we have not been able to give an analytical description of this overlapping regime.

\subsection{Numerical Verification}\label{Sec:numerics}
In this section, we will verify our analytical predictions by considering some benchmark models.  We will show that the asymptotic solution agrees well with the numerics.

We begin with the following model in five spacetime dimensions ($d=4$) inspired by supergravity theory.
\begin{equation}\label{5DStringModel}
\begin{split}
\mathcal{L}^{(5)}=-\frac{1}{2}(\partial_\mu\psi)^2-&\frac{\sinh^2(\kappa\psi/2)}{2}\left(\partial_\mu\theta-\frac{\sqrt{3}}{L}A_\mu\right)^2\\
&+\frac{3}{L^2}\cosh^2\frac{\psi}{2}(5-\cosh\psi)-\frac{1}{4}F_{\mu\nu}F^{\mu\nu}\,,
\end{split}
\end{equation}
with $\kappa$ a free constant. When $\kappa=2$, the theory can be lifted to a class of solutions of type IIB supergravity, based on D3-branes at the tip of a Calabi-Yau cone~\cite{Gubser:2009qm}.\footnote{Comparing to the form of~\cite{Gubser:2009qm}, the rescaling $A_\mu\rightarrow\frac{\sqrt{3}}{2L} A_\mu$ was made to have a standard normalization for the kinetic term of $U(1)$ sector. We also redefined $\eta=\psi$.} The coupling $\mathcal{F}=\sinh^2(\kappa\psi/2)/2\sim e^{\kappa\psi}$ for sufficient large value of $\psi$. Without loss of generality, we will set $L=1$ and will consider planar black holes. 

The resulting hairy black hole~\eqref{MetricFieldAnsatz} is asymptotically AdS and the expansion of matter fields near the AdS boundary $z=0$ is given by
\begin{equation}\label{UVform}
 \psi=\psi_s z+\dots+\psi_v z^3+\cdots,\quad
 A_t=\mu+\cdots-\frac{\rho}{2}z^2+\cdots\,,
\end{equation}
where $\psi_s$ is considered as the source of the scalar operator and $\psi_v$ as the expectation value from the viewpoint of the dual field theory. The two constants $\mu$ and $\rho$ correspond to the chemical potential and charge density, respectively. In the absence of source,  the development of scalar hair breaks the $U(1)$ symmetry spontaneously, mimicking the superconducting phase~\cite{Franco:2009yz,Kiritsis:2015hoa}. We shall focus on the planar black holes with $\psi_s=0$ and consider the grand canonical ensemble with $\mu=1$ unless explicitly stated otherwise. 
In the absence of scalar hair, the solution can be solved analytically, which is nothing but the AdS Reissner-Nordstr\"{o}m black hole.
As we will show, there is a critical temperature $T_c$ below which the scalar hair develops spontaneously. The scalar hair necessarily removes the inner horizon of AdS Reissner-Nordstr\"{o}m black hole.

\paragraph{Kasner inversion} Notice that $d=4$. The dynamics of $\widetilde{\alpha}$ for the Kasner inversion process now reads
\begin{equation}\label{5DInvAlphaDiff}
6z\widetilde\alpha\widetilde\alpha''-12z\widetilde\alpha'^2+\widetilde\alpha'\widetilde\alpha(\widetilde\alpha^2-6)=0\,,
\end{equation}
and the law describing the Kasner inversion is
\begin{equation}\label{5DKasInvLaw}
    \alpha \alpha_I=12\,.
\end{equation}

\paragraph{Kasner transiton} The differential equation satisfied by the Kasner transition process of $\widetilde{\alpha}$ becomes
\begin{equation}\label{5DTranAlphaDiff}
z \widetilde{\alpha}''+7\widetilde\alpha'\pm|\kappa|\widetilde\alpha\widetilde\alpha'=0\,,
\end{equation}
and the law for the Kasner transition is
\begin{equation}\label{5DKasTranLaw}
  \alpha +\alpha_T=\pm\frac{12}{|\kappa|}\,,
\end{equation}
with the plus and minus signs corresponding to $\alpha>0$ and $\alpha<0$, respectively. 

\begin{figure}[H]
\centering
\includegraphics[width=0.65\textwidth]{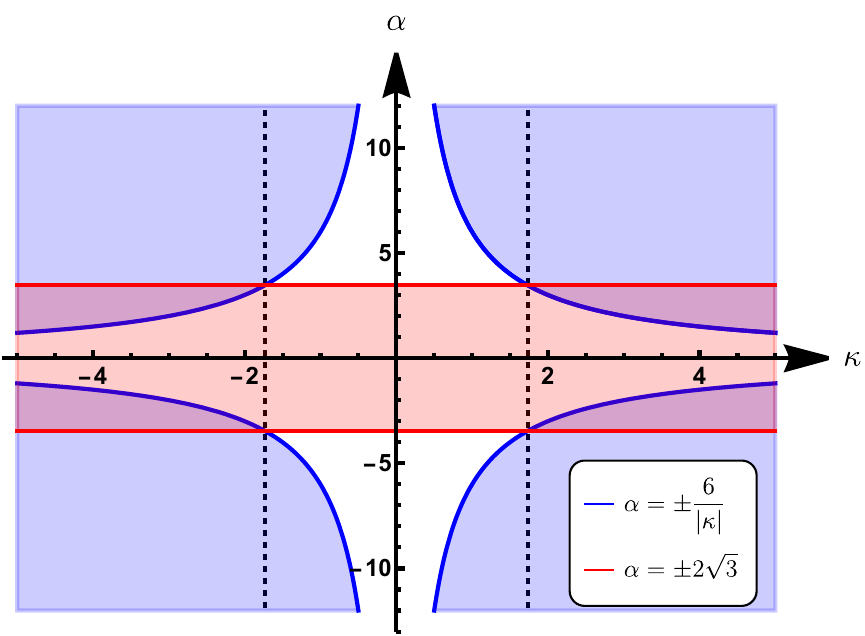}
\caption{The $\kappa$-$\alpha$ phase diagram for the benchmark model~\eqref{5DStringModel}.
The two vertical dashed lines at $\kappa=\pm\sqrt{3}$ divide the phase diagram into three parts. The middle part with $-\sqrt{3}<\kappa<\sqrt{3}$ corresponds to \textbf{Case I}, and the outer parts $|\kappa|>\sqrt{3}$ correspond to \textbf{Case III}. In addition, \textbf{Case II} is precisely given by the two vertical lines.
}\label{kappa_alpha_diagram}
\end{figure}

Based on the above results, we obtain the $\alpha$-$\kappa$ phase diagram presented in Fig.~\ref{kappa_alpha_diagram}.  For each case, a numerical example will be given.

\subsubsection{\texorpdfstring{$\kappa=2$}{TEXT}}
We first consider the case with $\kappa=2$, which is a top-down model that can be embedded into type IIB supergravity~\cite{Gubser:2009qm}. This model ($\kappa=2>\sqrt{3}$) belongs to \textbf{Case III}, \emph{i.e.} the outer parts of Fig.~\ref{kappa_alpha_diagram}. As the temperature is lowed, the scalar hair will develop spontaneously below the critical temperature $T_c=0.026\mu$, triggering a second order phase transition from the Reissner-Nordstr\"{o}m black hole to the charged hairy black hole, known as the holographic superconductor phase transition~\cite{Gubser:2009qm}.

\begin{figure}[htbp]
    \centering
    \begin{minipage}{0.495\textwidth}
        \centering
        \includegraphics[width=1\textwidth]{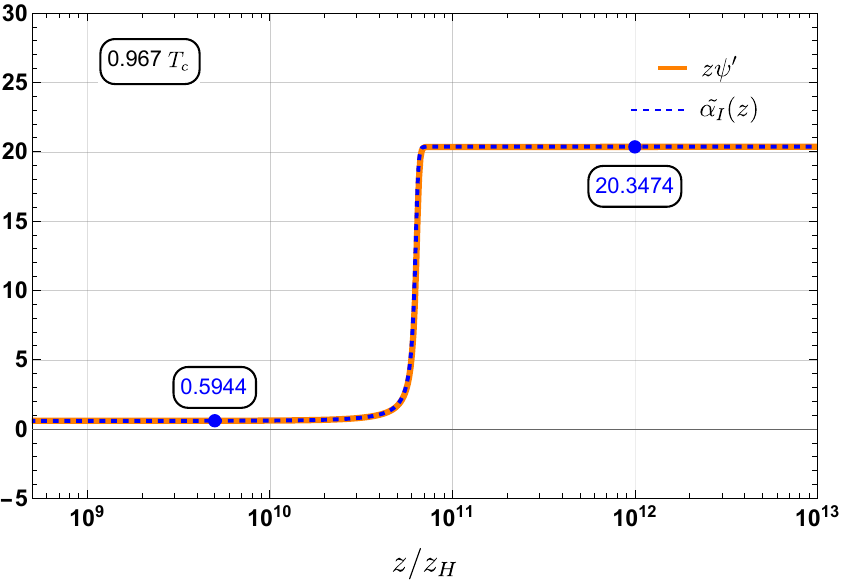}
    \end{minipage}
        \begin{minipage}{0.495\textwidth}
        \centering
        \includegraphics[width=1\textwidth]{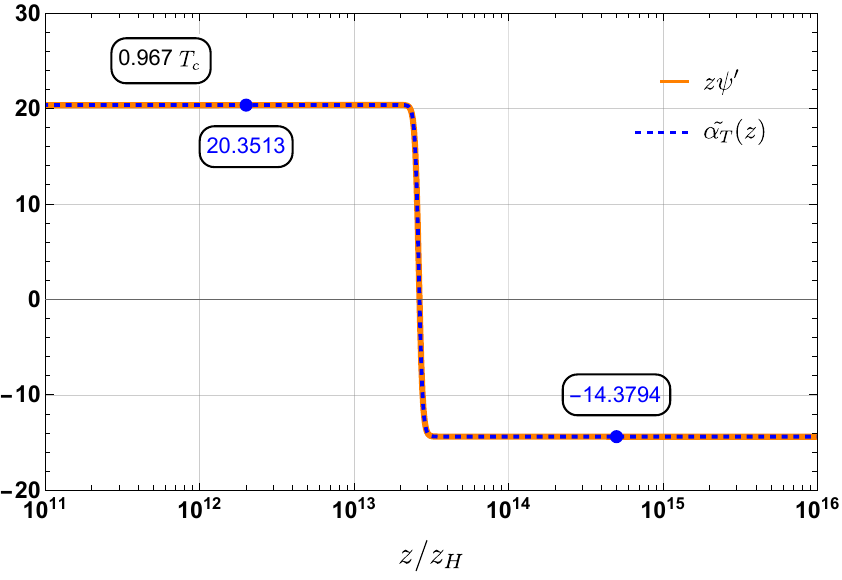}
    \end{minipage}
    \caption{A direct comparison of the analytical description~\eqref{5DTranAlphaDiff} (blue dashed curve) and the numerical one (solid orange curve) for \textbf{Kasner inversion} (left panel) and \textbf{Kasner transition} (right panel). Note from~\eqref{psiPresolve} that $\widetilde\alpha(z)=z\psi'(z)$. Each platform corresponds to a Kasner epoch with the number denoting the value of $\alpha$. We consider the hairy black hole at $T=0.967T_c$. The approximation~\eqref{5DTranAlphaDiff} is in excellent agreement to the profile from the full equations of motion $\eqref{EoM:psi}$-$\eqref{EoM:h}$. We have considered the model~\eqref{5DStringModel} with $\kappa=2$, \emph{i.e.} a top-down theory from supergravity~\cite{Gubser:2009qm}.}
    \label{5DTransition}
\end{figure}

In order to check whether the analytical description~\eqref{5DTranAlphaDiff} and~\eqref{5DInvAlphaDiff} can capture all the important effects describing the Kasner inversion and transition, we compare the profile of $\widetilde{\alpha}=z\psi'$ from the analytical one~\eqref{5DTranAlphaDiff} and~\eqref{5DInvAlphaDiff} with the numerical solution of the full equations of motion $\eqref{EoM:psi}$-$\eqref{EoM:h}$ in Fig.~\ref{5DTransition}. One can see that $\widetilde{\alpha}$ is almost a constant in each Kasner epoch, while it suffers from a sudden change at certain transformation points. We present the value of $\alpha$ in each Kasner epoch by fitting the numerical solutions (solid orange curve). There are two Kanser transformations presented in Fig.~\ref{5DTransition}, where the left panle is for a Kasner inversion and the right one for a Kasner transition. It is clear that our analytic approach is able to capture the key features of the Kasner transition. It gives an excellent description of transformation found numerically. Moreover, we have checked various numerical examples with a sequence of alternation of Kasner epochs, and find a good agreement with our transformation rule.

Since our top-down model pertains to \textbf{Case III}, the parameter space of Kasner inversion and the one of Kasner transition have an overlap. Outside the overlapping region $3<|\alpha|<2\sqrt{3}$, the alternation between adjacent Kasner epochs is described by the transformation laws~\eqref{5DKasInvLaw} and~\eqref{5DKasTranLaw}. Such standard transformation is clearly visible from Fig.~\ref{5DTransition}. In contrast,
when $\alpha$ falls into the overlapping region, the intricate interaction between the Kasner inversion and transition develops and the transformation behavior cannot be simply captured by~\eqref{5DKasInvLaw} and~\eqref{5DKasTranLaw}.

\begin{figure}[H]
\centering
\includegraphics[width=0.75\textwidth]{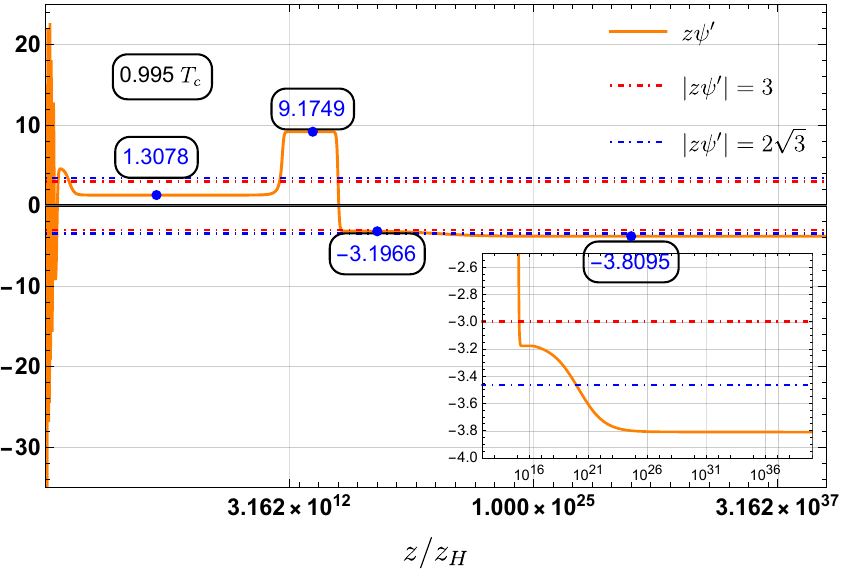}
\caption{The configuration of $z\psi'$ inside the hairy black hole at $T=0.995T_c$ for the model~\eqref{5DStringModel} with $\kappa=2$. The dashed red and blue curves mark $|z\psi'|=3$ and $|z\psi'|=2\sqrt{3}$, respectively. The value of $\alpha$ for each Kanser epoch is given explicitly. When $z\psi'=-3.1966\in(-2\sqrt{3}, -3)$, it goes through a competitive process that can't be described by our inversion or transition law. The inset zooms in on this transformation. After this process, the system arrivals at a Kanser epoch with $\alpha=-3.8095$. The present model can be embedded into supergravity~\cite{Gubser:2009qm}.}\label{5DAlternateCase3}
\end{figure}

As shown in Fig.~\ref{5DAlternateCase3} for $T=0.995T_c$, after the ER collapse and scalar oscillation, one has a Kanser epoch with $\alpha=1.3078$. Then after a Kasner inversion and a transition, the resulting epoch has $\alpha=-3.1966$ which is within the overlapping region $(-2\sqrt{3},-3)$. The profile of $\psi$ becomes no more logarithmic and the value of $\alpha=z\psi'$ decreases towards the interior in the present case.  By a complicated transformation process, it arrivals at a Kanser epoch with $\alpha=-3.8095$ that deviates significantly from the value predicted by Kasner inversion or transition. Since $\alpha$ falls out the overlapping zone, more standard Kasner alternations will be triggered.

\subsubsection{\texorpdfstring{$\kappa=\sqrt{3}$}{TEXT}}
Then, we choose $\mathcal{F}={\sinh^2(\sqrt{3}\psi/2)}/2$ of model~\eqref{5DStringModel}, which yields $\kappa=\sqrt{3}$ and thus belongs to \textbf{Case II} (the vertical dashed line of Fig.~\ref{kappa_alpha_diagram}). As we have mentioned, this is similar to the four dimensional top-down model studied in~\cite{An:2022lvo}. Unless $\alpha=\pm 2\sqrt{3}$, there will be generically a never-ending chaotic alternation of Kasner epochs towards the singularity.

\begin{figure}[H]
\centering
\includegraphics[width=0.75\textwidth]{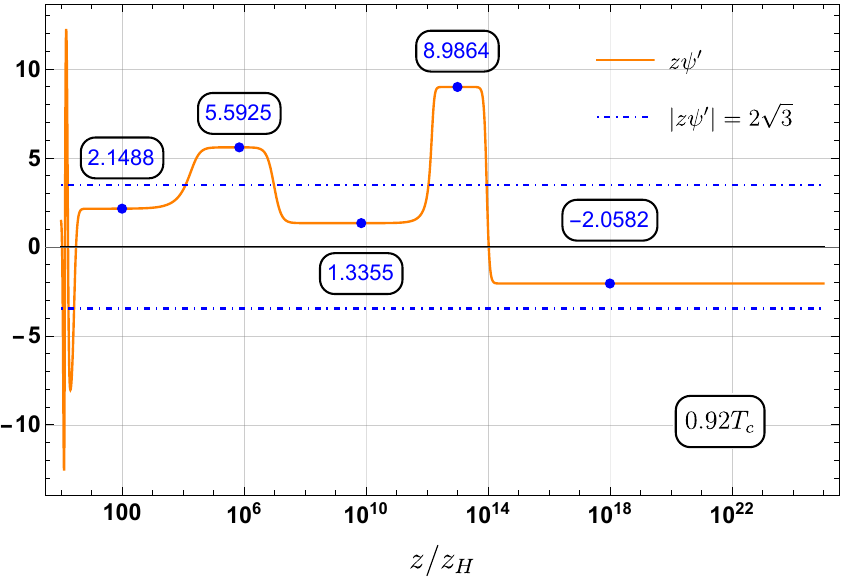}
\caption{The interior configuration of $z\psi'$ at $T=0.92T_c$ for the model~\eqref{5DStringModel} with $\kappa=\sqrt{3}$. Both the boundaries of the Kasner inversion and transition are at $|\alpha|=2\sqrt{3}$, so it will be an infinite Kasner alternation process. The value of $\alpha$ for each Kasner epoch is given by solving the full equations of motion. The validity of the transformation rule~\eqref{ruleII} for the alternation of Kasner epochs is manifest.}\label{5DAlternateCase2}
\end{figure}

In Fig.~\ref{5DAlternateCase2}, we show $z\psi'$ as a function of $z$ inside the hairy black hole for $T=0.92 T_c$ with $T_c=0.015\mu$. Two Kasner inversions and two Kasner transitions are manifest. It is easy to check that the value of $\alpha$ for each Kasner epoch agrees with our transformation rule
\begin{equation}\label{ruleII}
\begin{split}
\alpha \alpha_I=12,\quad |\alpha|<2\sqrt{3}\,,\\
\alpha+\alpha_T=\pm 4\sqrt{3}, \quad |\alpha|>2\sqrt{3}\,.
\end{split}
\end{equation}
For example, the first Kasner epoch is around $z/z_H=100$ with $\alpha_1=2.1488$. Thus, a Kasner inversion is triggered and results in a Kasner eopch with $\alpha_2=5.5925$. One finds $\alpha_1 \alpha_2=12.0172$ as predicted by our analytic approach. Since $\alpha_2>2\sqrt{3}$, there should be the third Kasner epoch via the Kasner transition. We find that $\alpha_3=1.3355$ and thus $\alpha_2+\alpha_3=6.928$, in good agreement with Kasner transition law~\eqref{ruleII}.  The new Kasner regime is again unstable and the process would repeat itself for ever.

\subsubsection{\texorpdfstring{$\kappa=3/2$}{TEXT}}
We can also realize the situation for \textbf{Case I} by, for example, setting $\kappa=3/2$, \emph{i.e.} $\mathcal{F}(\psi)={\sinh^2(3\psi/4)}/2$. It exists a stable region $2\sqrt{3}<|\alpha|<4$ as visible from Fig.~\ref{kappa_alpha_diagram}. Once a Kasner epoch falls into this region, it will stay at this Kasner epoch towards the singularity without suffering from further Kasner alternation.

\begin{figure}[H]
\centering
\includegraphics[width=0.75\textwidth]{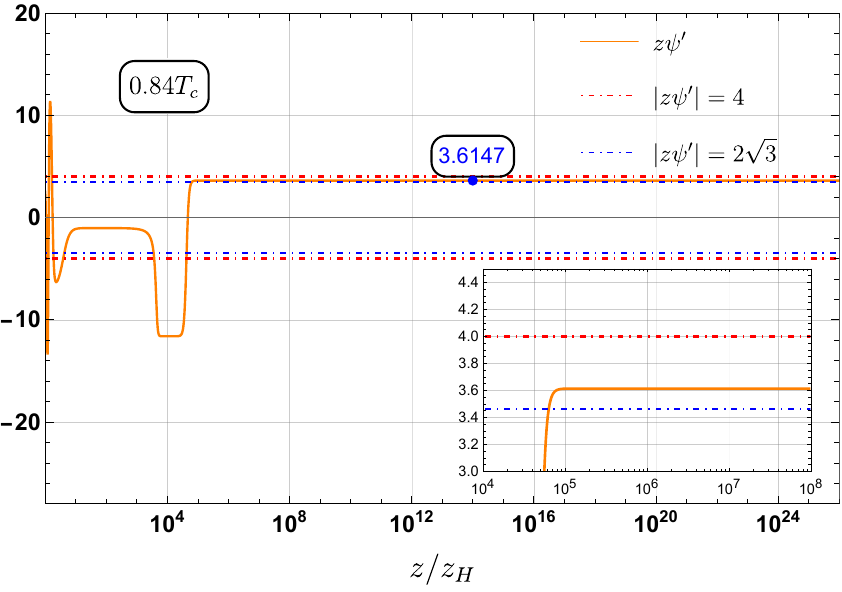}
\caption{The interior profile for $z\psi'$ for the model~\eqref{5DStringModel} with $\kappa=3/2$ and $T=0.84T_c$. There is a stable region with $2\sqrt{3}<|\alpha|<4$. One finds a stable Kasner epoch with $\alpha=3.6147$ for $z/z_H>10^5$. The inset zooms in on the transition.}\label{AlternateCase1}
\end{figure}

A typical example is given in Fig.~\ref{AlternateCase1} for the hairy black hole at $T=0.84 T_c$ where the critical temperature $T_c=0.007\mu$. After a Kasner transition around $z/z_H=10^5$, the system jumps to the Kasner epoch with $\alpha=3.6147$ within the stable region. Due to the limitation of computing power, we are not able to solve the full equation of motion $\eqref{EoM:psi}$-$\eqref{EoM:h}$ for sufficiently large $z$.
Nevertheless, the sufficiently long stable phenomenon that appears in Fig.~\ref{AlternateCase1} can be considered as a strong support.

\section{Interior Dynamics for General Coupling and Potential}\label{Sec:general}
We have obtained two typical transformations at later interior times, \emph{i.e.} the Kasner inversion caused by the kinetic energy term of the gauge field and the Kasner transition dominated by the exponential coupling $\mathcal{F}\sim e^{\kappa\psi}$. In both cases, the scalar potential $V(\psi)$ is negligible, and the transformation law can be given analytically. It is challenging to understand other cases due to the highly nonlinear nature of the system. However, from a large number of numerical examples, we find that there exist Kasner structures and Kasner transformations for general coupling $\mathcal{F}$ and potential $V$ under certain condition. In this section we aim at providing general features.

\subsection{Case with General Coupling $\mathcal{F}$}
We begin with the simplest case for which the contribution of $V$ is neglected. We also note that $h'$ is integrable, thus the order of $h'/h$ is smaller than $1/z$ and can be neglected in~\eqref{EoM:psi}. At this time, the approximate equation of motion about $\psi$ is given by
\begin{equation}\label{FDominates}
\psi''=-\frac{1}{z}\psi'-\frac{q^2A_t^2}{z^{2d}h^2}\frac{\mathrm{d}\mathcal{F}}{\mathrm{d}\psi}\,.
\end{equation}
We can obtain from the above equation that
\begin{equation}\label{feedF}
\widetilde\alpha'=-\frac{q^2A_t^2}{z^{2d-1}h^2}\frac{\mathrm{d}\mathcal{F}}{\mathrm{d}\psi}\quad\Rightarrow\quad\widetilde\alpha'\frac{\mathrm{d}\mathcal{F}}{\mathrm{d}\psi}=-\frac{q^2A_t^2}{z^{2d-1}h^2}\left(\frac{\mathrm{d}\mathcal{F}}{\mathrm{d}\psi}\right)^2<0\,,
\end{equation}
where we have used~\eqref{psiPresolve}.
Hence, for the case with positive $\frac{\mathrm{d}\mathcal{F}}{\mathrm{d}\psi}$, the value of $\widetilde\alpha$ will decrease towards the deep interior, while the negative $\frac{\mathrm{d}\mathcal{F}}{\mathrm{d}\psi}$ will result in the increase of $\widetilde\alpha$. 

For the exponential coupling $\mathcal{F}\sim e^{\kappa\psi}$ discussed in Subsection~\ref{SubSec:KasTrans}, when $\kappa$ is within~\eqref{LowTransLimit}, the coupling $\mathcal{F}$ dominates the system and leads to a transformation to a stable Kasner regime with a smaller value of $|\alpha|$ via the Kasner transition. For the coupling function with super-exponential and more general forms, it is difficult to analytically obtain the transformation process. Nevertheless, due to the similar  mechanism discussed at the end of Subsection~\ref{subsec:inversion}, some generic features can be given. Suppose there is a transformation from a Kasner epoch to another one. For $\frac{\mathrm{d}\mathcal{F}}{\mathrm{d}\psi}>0$, the new Kasner epoch will have a smaller value of the Kasner exponent $\alpha$, while  it will have a larger Kasner component for $\frac{\mathrm{d}\mathcal{F}}{\mathrm{d}\psi}<0$. Such transformation process could result in a sequence of Kasner epochs until the contribution from $\mathcal{F}$ term becomes negligible and the system settles down to a stable epoch.\,\footnote{Here we assume that each epoch at late times has a Kasner form. It is possible that there develops no Kasner epoch for some choice of $\mathcal{F}$.}

We now give an example for the interior dynamics of $\widetilde\alpha$. In order to highlight the role of $\mathcal{F}$, we consider the five dimensional model with the super-exponential coupling.
\begin{equation}\label{modelSF}
\mathcal{F}=\sinh(\sinh^2(\psi)),\quad V=-12-\frac{3}{2}\psi^2, \quad q=\sqrt{3}\,.
\end{equation}
One has $\mathcal{F}\sim \text{exp}(e^{2|\psi|})$ asymptotically for large $\psi$. Ignoring the scalar potential, the equation of motion about $\psi$ is approximated by
\begin{equation}\label{AppEoMPsiForF}
\alpha'\simeq-\frac{h'}{h}\alpha-\frac{3A_t^2}{z^{7}h^2}\frac{\mathrm{d}\mathcal{F}}{\mathrm{d}\psi}\,,
\end{equation}
where we have also included the contribution from $h'/h$, since the non-integrability of $h'$ will also cause the instability of a Kasner epoch when its exponent $|\alpha|<2\sqrt{3}$ (see Subsection~\ref{subsec:inversion}).

\begin{figure}[H]
\centering
\includegraphics[width=0.7\textwidth]{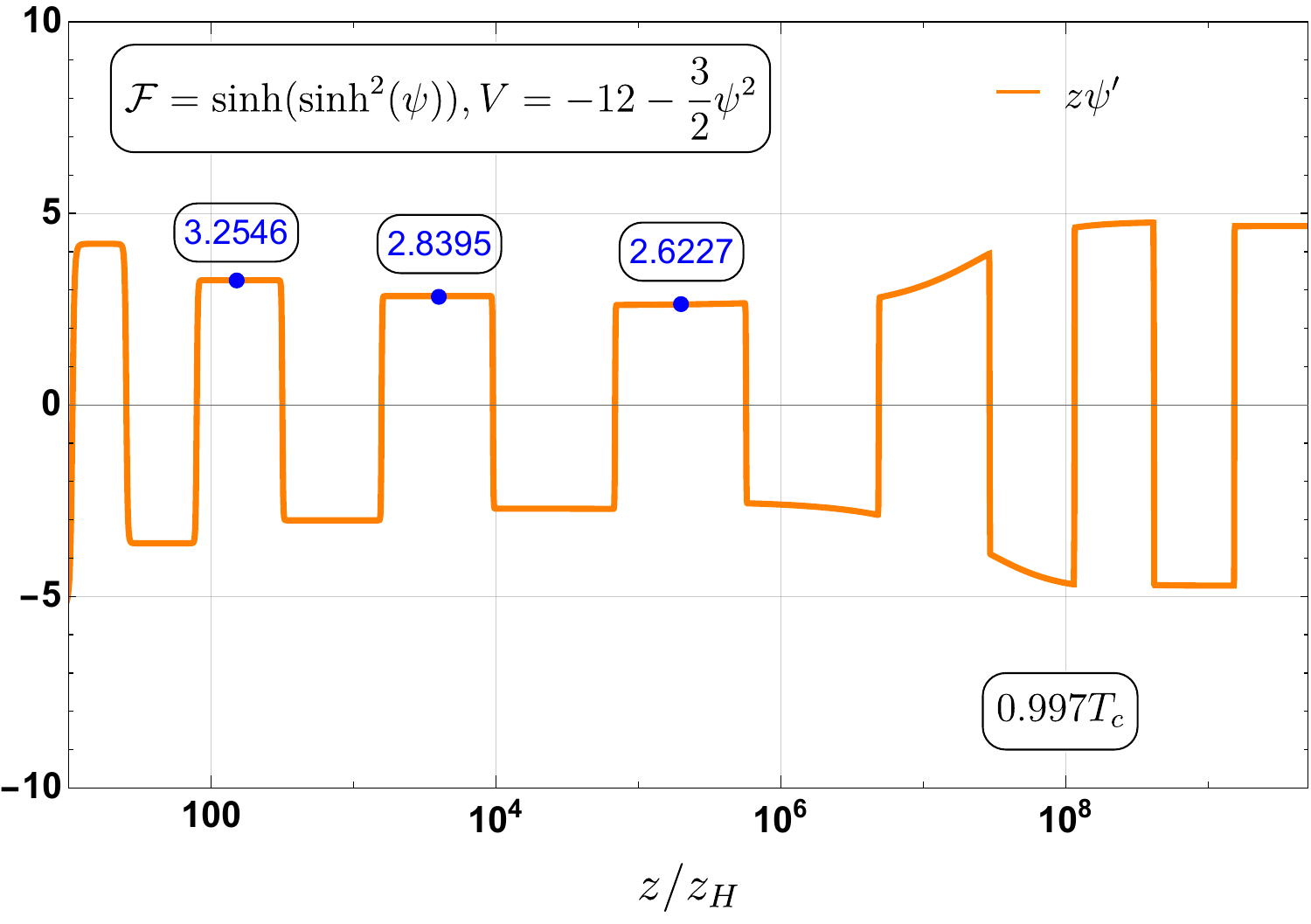}
\caption{Evolution of $z\psi'$ as a function of $z$ behind the event horizon $z_H$. We consider the planar hairy solution at $T=0.997T_c$ for the model~\eqref{modelSF} with a super-exponential coupling $\mathcal{F}$. Each platform corresponds to a Kasner epoch with a constant Kasner exponent $\alpha$. The value of $\alpha$ is labelled in some Kanser epochs.
There develops a sequence of Kasner epochs as well as non-Kasner epochs.}\label{SuperF}
\end{figure}

\begin{figure}[H]
\centering
\includegraphics[width=0.49\textwidth]{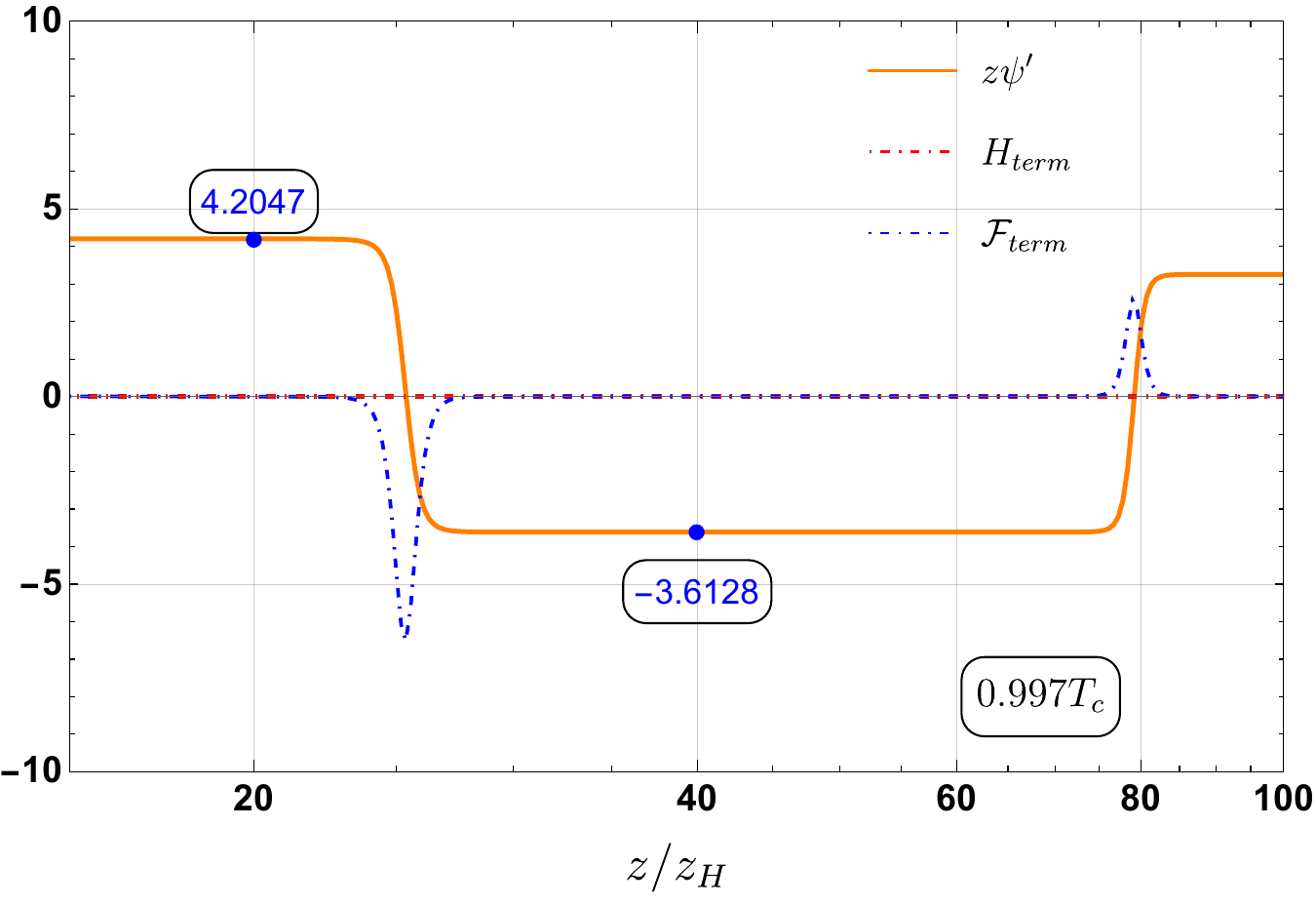}
\includegraphics[width=0.49\textwidth]{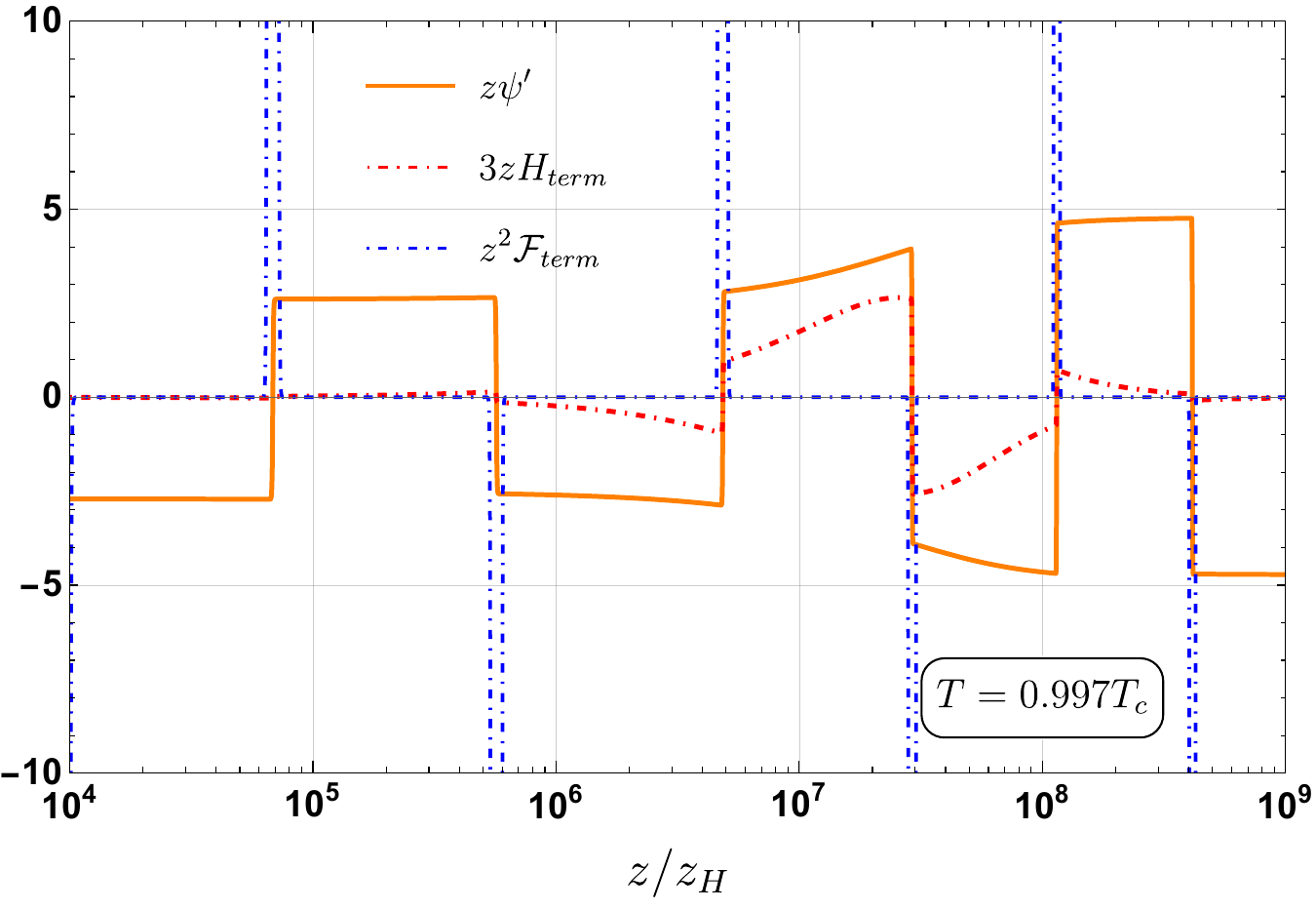}
\caption{Zoom in on the evolution of $z\psi'$ in Fig.~\ref{SuperF}. The interior behavior is dominated by the two terms in the right hand of~\eqref{AppEoMPsiForF}, for which 
the first term $\mathcal{H}_{\text{term}}=-\frac{h'}{h}\alpha$ is denoted by the red dashed curve and the second term $\mathcal{F}_{\text{term}}=-\frac{3A_t^2}{z^{7}h^2}\frac{\mathrm{d}\mathcal{F}}{\mathrm{d}\psi}$ is denoted by the blue dashed curve. The left panel shows two Kasner transformations dominated by $\mathcal{F}_{\text{term}}$. When  $10^6<z/z_H<10^8$, in the right panel non-Kasner epochs are manifest where both terms come into play.}\label{SuperF_FHterm}
\end{figure}

The interior evolution of $\widetilde\alpha=z\psi'$ is presented in Fig.~\ref{SuperF}, which exhibits a sequence of Kasner epochs as well as some non-Kasner regions. The two Kasner epochs at the left-most position in Fig.~\ref{SuperF} has $\alpha_1=4.2047$ and $\alpha_2=-3.6128$ (as visible from the left panel of Fig.~\ref{SuperF_FHterm}). Both are outside the non-integrability condition~\eqref{KasInvCon} and thus $h'/h$ would not play a dominant role. It is manifest from the left panel of Fig.~\ref{SuperF_FHterm} that the Kasner alternation is triggered by the $\mathcal{F}$ term, thus does not obey the Kasner inversion law~\eqref{KasInvLaw}. As expected, the transformation results in a Kasner epoch with a smaller $\alpha_2$. As a negative $\alpha_2$ will result in a negative $\frac{\mathrm{d}\mathcal{F}}{\mathrm{d}\psi}$ as the interior time evolves, a new Kasner epoch with a larger exponent $\alpha$ is anticipated. One can see from Fig.~\ref{SuperF} that the third Kasner exponent $\alpha_3=3.2546>\alpha_2$.

Since $\alpha_3=3.2546<2\sqrt{3}$, the $h'/h$ term itself can trigger the instability of third Kasner epoch at late interior times. We have a sequence of Kasner alternations until $z/z_H=10^5$ with the amplitude of $\alpha$ decreasing. As visible from the right panel of Fig.~\ref{SuperF_FHterm}, the $h'/h$ term becomes important to the dynamics. In the absence of $\mathcal{F}$ term, one expects to have a Kasner inversion. Since both terms can not be ignored, we find some alternations between non-Kasner epochs when $10^6<z/z_H<10^8$. For much larger $z/z_H$, the system shows more Kasner epochs.  We anticipate an infinite sequence of epochs, as the scalar field rolls back and forth in its coupling $\mathcal{F}$.

\subsection{Case with General Potential $V$}

So far we have required that the contribution from the scalar potential $V(\psi)$ should be ignored. This allows $V$ to be arbitrary algebraic functions, including polynomial functions. However, for $V$ that diverges faster than the exponential growth, the scalar potential usually comes into play and our transformation laws~\eqref{KasInvLaw} and~\eqref{KasTranLaw} will be invalid. For example, an even super-exponential potential has been shown to trigger an infinite number of bounces for Kasner epochs~\cite{Cai:2020wrp,Hartnoll:2022rdv}. In particular, by assuming the rate of growth is approximately constant over the bounce ($|V'''V'/V''^2|\ll 1$), an analytic expression for the bounces between each Kasner epoch have been discussed in~\cite{Hartnoll:2022rdv}. It was shown that at late interior times the Kasner exponent $\alpha$ tends to zero and the interior metric slowly approaches the Schwarzschild singularity. Nevertheless, the analysis of~\cite{Hartnoll:2022rdv} does not apply to general cases.

When the scalar potential dominates, the approximate equation of motion about $\psi$ is given by
\begin{equation}\label{VDominatesPsi}
\psi''=-\frac{1}{z}\psi'+\frac{\mathrm{e}^{-\chi/2}}{z^{d+2}h}\frac{\mathrm{d}V}{\mathrm{d}\psi}\,,
\end{equation}
from which we get 
\begin{equation}\label{alphaV}
\widetilde\alpha'=\frac{\mathrm{e}^{-\chi/2}}{z^{d+1}h}\frac{\mathrm{d}V}{\mathrm{d}\psi}\quad\Rightarrow\quad \widetilde\alpha'\frac{\mathrm{d}V}{\mathrm{d}\psi}=\frac{\mathrm{e}^{-\chi/2}}{z^{d+1}h}\left(\frac{\mathrm{d}V}{\mathrm{d}\psi}\right)^2<0\,,
\end{equation}
since $h<0$ inside the hairy black hole. One immediately finds a similar feature as the previous case, see~\eqref{feedF}. The value of $\widetilde\alpha$ known as Kasner velocity in~\cite{Hartnoll:2022rdv} will decrease towards the deep interior for $\frac{\mathrm{d}V}{\mathrm{d}\psi}>0$, while it will increase for $\frac{\mathrm{d}V}{\mathrm{d}\psi}<0$.

\begin{figure}[H]
\centering
\includegraphics[width=0.49\textwidth]{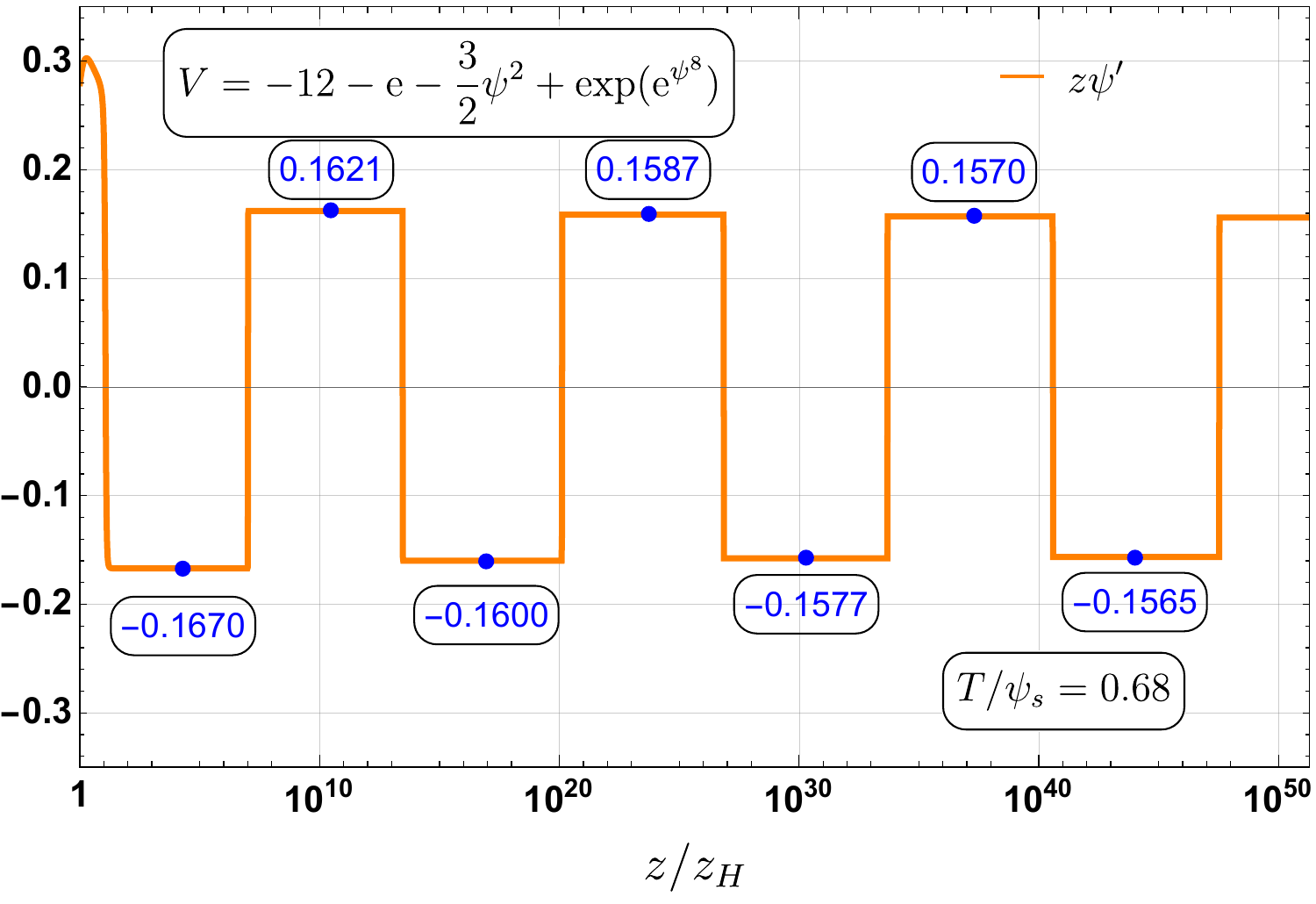}
\includegraphics[width=0.50\textwidth]{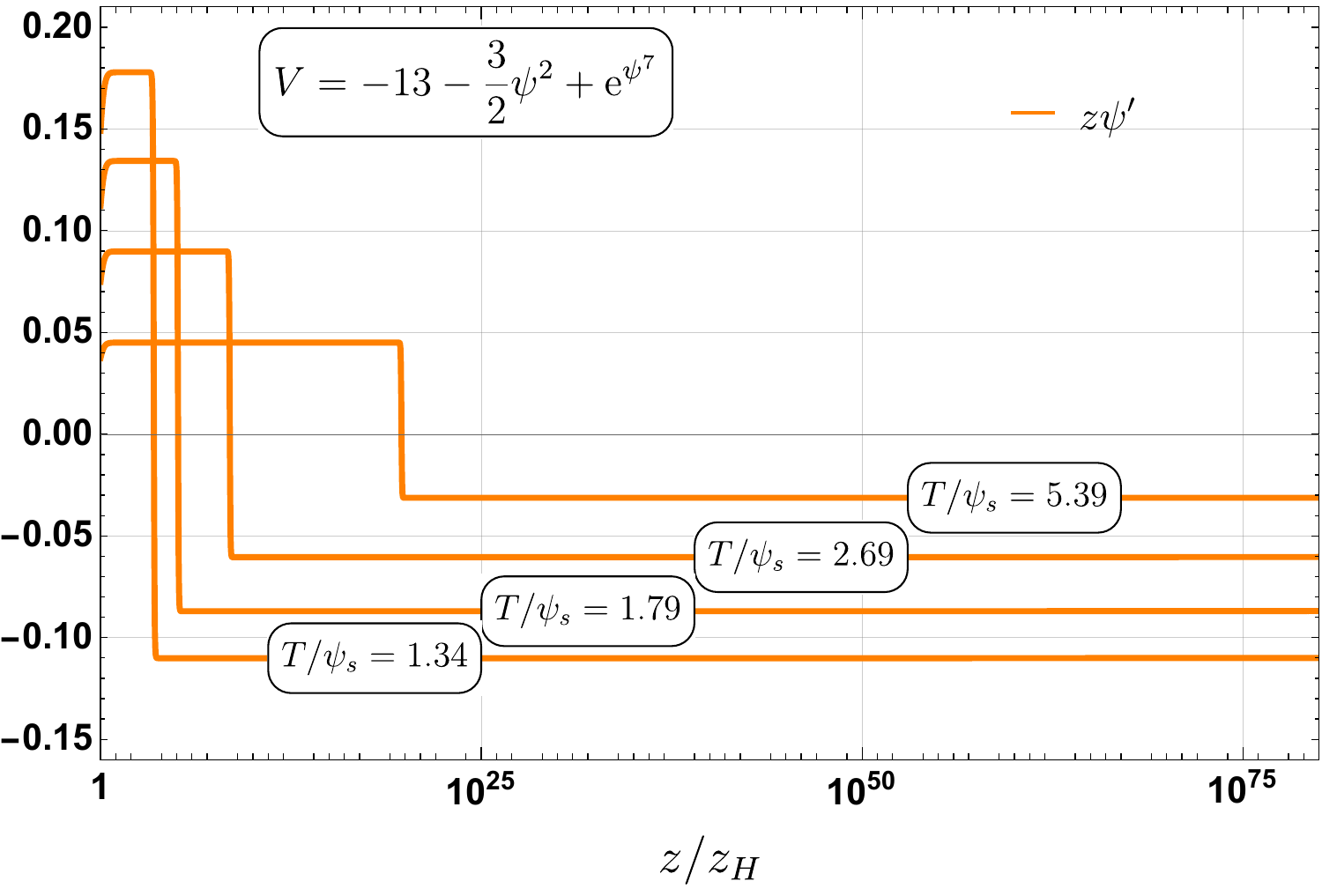}
\caption{Kasner structure and transformation triggered by super-exponential potentials for the Einstein-scalar theory. The left panel is dominated by an even super-exponential potential $V\sim \text{exp}(\mathrm{e}^{\psi^{8}})$ and the right one is dominated by $V\sim \mathrm{e}^{\psi^{7}}$. To highlight the role of scalar potential, we turn off the $U(1)$ gauge field. The scalar potentials are chosen to have the asymptotic behavior as $\psi\rightarrow 0$ near the AdS boundary $V=-12-\frac{3}{2}\psi^2+\cdots$, for which the boundary expansion is given by~\eqref{UVform}. To obtain the hairy black holes in such charged neutral case, we fix the boundary source for the scalar $\psi_s=1$. We consider the planar horizon case in five dimensional spacetime.}\label{SuperVKasner}
\end{figure}

For an even super-exponential potential $V\sim e^{\psi^{2n}}$ with $n$ a positive integer, its derivative to $\psi$ is an odd function. We begin with a Kasner epoch with a positive Kasner exponent $\alpha_0$ and a positive $\psi$ (\emph{i.e.} $\frac{\mathrm{d}V}{\mathrm{d}\psi}>0$). The Kasner transformation triggered by the scalar potential, if it happens, will lead to a new Kasner epoch with a smaller Kasner exponent $\alpha_1<\alpha_0$. If $\alpha_1<0$, $\psi\sim\alpha_1\ln(z)$ will typically become negative towards interior and thus $\frac{\mathrm{d}V}{\mathrm{d}\psi}<0$. Then the scalar potential could trigger another Kasner transformation, giving the third Kasner epoch with a larger Kasner exponent $\alpha_2>\alpha_1$. Once $\alpha_2>0$, there would be the third Kasner transition to the epoch with a negative Kasner exponent, and so on. Thus, we could have an infinite sequence of Kasner epochs. This is what has been observed in the literature, see \emph{e.g.} Figure 1 of~\cite{Hartnoll:2022rdv}. Similar features can be found for more exotic potentials, see the left panel of Fig.~\ref{SuperVKasner} for $V\sim \exp({e^{\psi^8}})$.

On the other hand, for an super-exponential potential $V\sim e^{\psi^{2n+1}}$ with odd power, one has $\frac{\mathrm{d}V}{\mathrm{d}\psi}\sim \psi^{2n}e^{\psi^{2n+1}}$. For a Kasner epoch with $\alpha_0>0$ and $\psi>0$, one has $\frac{\mathrm{d}V}{\mathrm{d}\psi}>0$. The Kasner transformation triggered by the scalar potential, if it happens, will result in a new Kasner epoch with a smaller Kasner exponent $\alpha_1<\alpha_0$. Then $\psi$ will decrease and could even become negative at large $z$. In contrast to the even super-exponential case, the decrease of $\psi$ makes the contribution from scalar potential less important. In particular, once $\psi$ becomes sufficiently negative, $\frac{\mathrm{d}V}{\mathrm{d}\psi}\sim \psi^{2n}e^{\psi^{2n+1}}$ is suppressed super-exponentially. Thus, the system could settle down to a stable Kasner epoch, instead of experiencing an infinite sequence of Kasner epochs.
An example with $V\sim e^{\psi^{7}}$ is shown in the right panel of Fig.~\ref{SuperVKasner}. For each temperature, one can see a sufficiently long Kasner epoch with a negative Kasner exponent. Similar discussion applies to the case with the opposite sign, \emph{i.e.} $V\sim e^{-\psi^{2n+1}}$, for which the system would settle down to a Kanser epoch with a positive Kasner exponent $\alpha$. However, if one turns on both kinds of odd super-exponential potentials, there will exhibits an infinite sequence of Kasner epochs, as the scalar field rolls back and forth in such potential. 

For the sake of simplicity, we have turned off the $U(1)$ gauge potential $A_t$ in our numerical examples in Fig.~\ref{SuperVKasner}. Actually, the above discussion applies to the charged black holes without inner horizon, no matter $\mathcal{F}(\psi)$ is zero or not. In the absence of $\mathcal{F}$, one can not remove the inner horizon completely. But a neutral scalar generically leads to a black hole with no inner horizon~\cite{Hartnoll:2020rwq}\,\footnote{For such kind of hairy black holes, inner horizons do exist at some specific temperatures~\cite{Hartnoll:2020rwq}.}. In the presence of $A_t$, in order to have a finally stable Kasner epoch, the value of its Kasner exponent $\alpha$ should be outside~\eqref{KasInvCon}. Otherwise, a Kasner inversion in Subsection~\ref{subsec:inversion} will occur and the system will jump into a Kasner epoch with $\alpha$ outside~\eqref{KasInvCon}. If no other terms come into play, it will be the finally stable Kasner epoch.

\begin{figure}[H]
\centering
\includegraphics[width=0.49\textwidth]{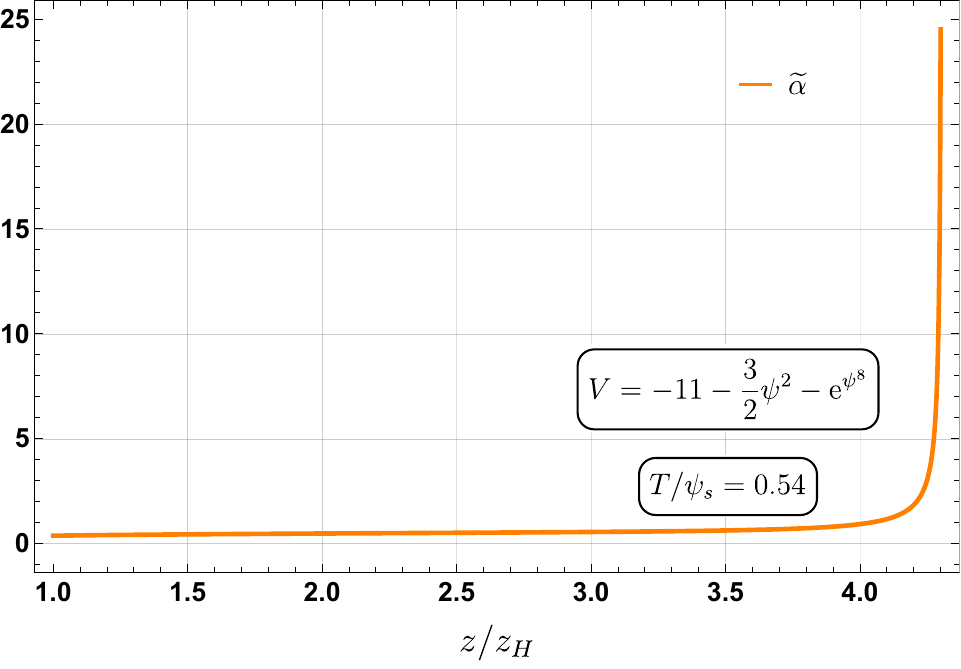}
\includegraphics[width=0.49\textwidth]{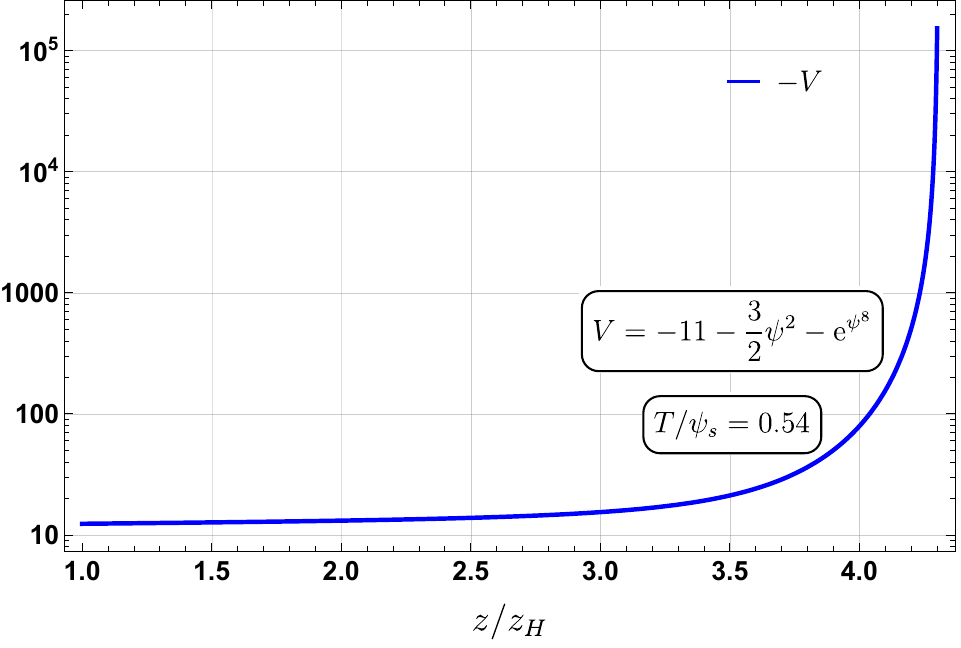}
\caption{Evolution of $\widetilde\alpha=z\psi'$ (left panel) and the scalar potential $V=-11-\frac{3}{2}\psi^2-e^{\psi^8}$ (right panel) as a function $z$. Both suffer a catastrophic increase above $z/z_H=4.2$ and there is no any Kasner epoch. The planar hairy black holes are numerically constructed with the gauge potential $A_t=0$ and the scalar source $\psi_s=1$.}\label{SuperVBreak}
\end{figure}

One key difference compared with the coupling term is that $\mathcal{F}$ is bound from below to ensure positivity of the kinetic term, while there is in principle no bound for $V(\psi)$. Thus the scalar potential could result in much richer interior behaviors. For example, we consider a negative even super-exponential potential $V\sim- e^{\psi^{2n}}$ with its derivative $\frac{\mathrm{d}V}{\mathrm{d}\psi}\sim -\psi^{2n-1}e^{\psi^{2n}}$. Therefore, choosing a point $z_i$ at which $\psi(z_i)>0$ and $\psi'(z_i)>0$, one has $\frac{\mathrm{d}V}{\mathrm{d}\psi}<0$. Then $\widetilde\alpha$ will increase as $z$ increases according to~\eqref{alphaV}. Therefore, the scalar will increase monotonically and the super-exponential potential would increase very quickly. If no other terms that can offset this catastrophic increase, the amplitude of $V$ will increase until reaching the singularity at which the spacetime terminates. We choose a strong potential that yields a rapidly increase of $V$ and $\widetilde\alpha$ in the interior, see Fig.~\ref{SuperVBreak}.
One can see from the left panel, there is no any Kasner epoch and the value of $\widetilde\alpha$ suffers a catastrophic increase above $z/z_H\simeq4.2$. The value of $V$ versus $z$ is shown in the right panel. So far, we are not able to show whether the scalar field could escape to infinity at a finite $z$.

\subsection{Case with both $\mathcal{F}$ and $V$}
According to our discussion in the previous two Subsections, once both $\mathcal{F}$ and $V$ are included and become important, the synergy and competition will result in very complicated interior dynamics. 

For illustration, we consider a model equipped with a super-exponential potential in five spacetime dimensions.
\begin{equation}\label{SuperVcase2}
V(\psi)=-13-\frac{3}{2}{\psi}^2+\mathrm{e}^{{\psi}^4},\quad \mathcal{F}=\sinh^2(\psi),\quad q=\sqrt{3}\,.
\end{equation}
It follows the same asymptotic behavior as~\eqref{UVform} at the AdS boundary. We focus on the planar hairy black holes for which the scalar hair develop spontaneously below the critical temperature $T_c=0.1221\mu$. 

\begin{figure}[H]
\centering
\includegraphics[width=0.49\textwidth]{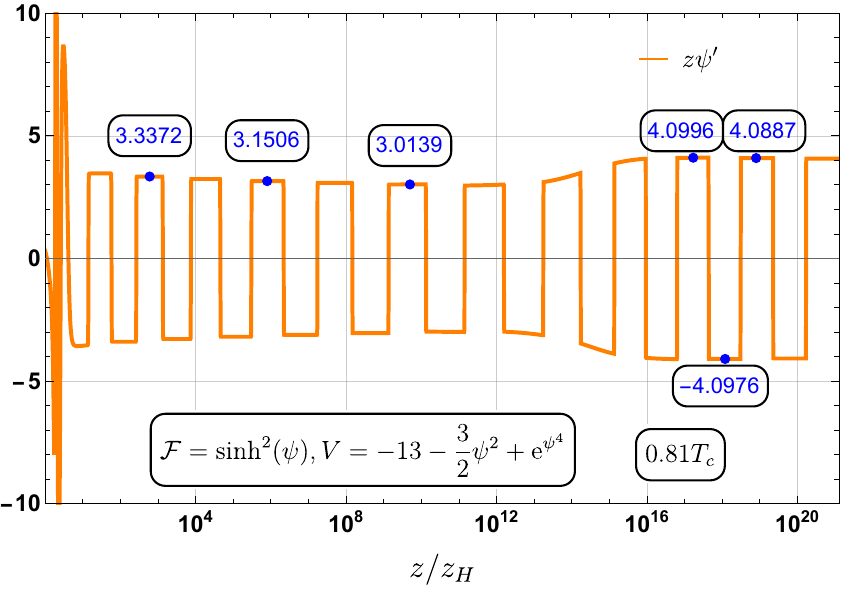}
\includegraphics[width=0.49\textwidth]{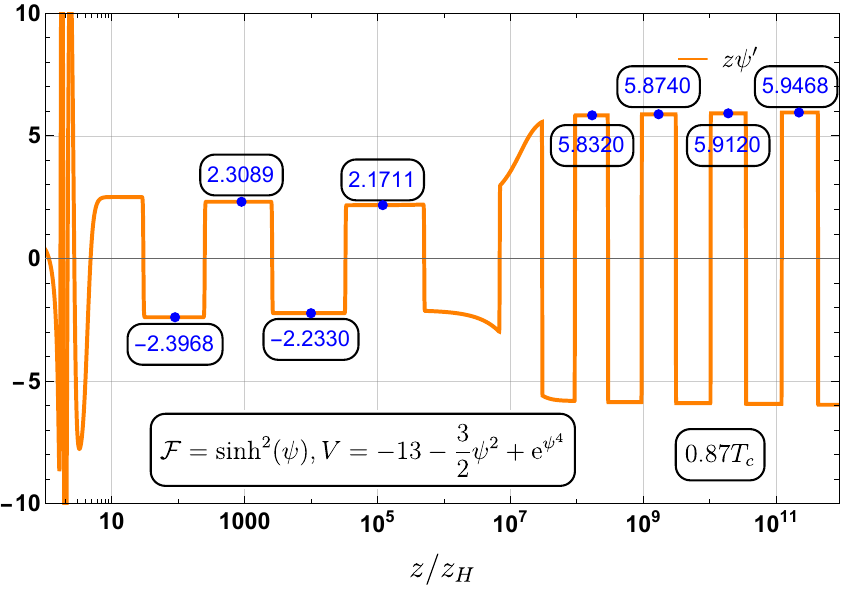}
\caption{Interior dynamics of the planar hairy black holes at $T=0.81T_c$ (left) and $T=0.87T_c$ (right) for the model~\eqref{SuperVcase2}. There develop complicated behaviors, including the presence of non-Kasner epochs and the random change of the amplitude of the Kasner exponent.}\label{FSinhVExp}
\end{figure}

The evolution of $\widetilde\alpha$ is presented in Fig.~\ref{FSinhVExp} from which more involved behaviors are manifest. One can see some sequences of Kasner epochs separated by abrupt bounces in which the Kasner exponent $\alpha$ changes sign. The amplitude of Kasner exponent $\alpha$ can increase or decrease. There also develop alternations between non-Kasner epochs, similar to what we have seen in Fig.~\ref{SuperF}. An analytic understanding of those interior dynamics is not yet available.

\section{Conclusion and Discussion}\label{Sec:ConAndDis}
We have studied the interior of hairy black holes in Einstein-Maxwell-scalar theory, which covers a large class of models considered in a recent body of research. It allows a Kanser universe characterized by the Kasner exponent $\alpha=z\psi'$ if the contributions from most of interactions to the interior dynamics can be dropped off, see Subsection~\ref{subKasner}. Kasner spacetimes continue to play a central role once more terms enter into the dynamics, but they may not persist for ever. We have been able to characterize the late interior time behaviors rather explicitly. 

We have uncovered two kinds of alternation of Kasner epoch. One is called Kasner inversion triggered by the non-integrability of $h'=(z^{-d} e^{-\chi/2}f)'$ discussed in Subsection~\ref{subsec:inversion}. More precisely, when $|\alpha|>\sqrt{2(d-1)(d-2)}$, the Kasner inversion yields a stable Kasner epoch if no additional term comes into play, and the two Kasner exponents before and after the Kasner inversion satisfy~\eqref{KasInvLaw}. In addition, when the coupling function $\mathcal{F}$ takes an exponential form $\mathcal{F}\sim e^{\kappa\psi}$ asymptotically at large $\psi$, the Kasner transition process will be triggered when $\kappa\alpha>(2d-2)$, see Subsection~\ref{SubSec:KasTrans}. The transformation law of the Kasner transition between two adjacent Kasner epochs is given by~\eqref{KasTranLaw}. Depending on the spacetime dimension and the coupling constant $\kappa$, we have predicted three different cases of Kasner alternation at later interior times (see Fig.~\ref{fig:KasnerTI}). Our analytical expressions have been corroborated by numerical solutions to the full equations of motion~$\eqref{EoM:psi}$-$\eqref{EoM:h}$. Several models have been checked in Section~\ref{Sec:numerics}, including a top-down model from supergravity.

In the case of the BKL limit, the billiard model can well describe the motion of space-time and fields near the singularity for scalarized black holes~\cite{Henneaux:2022ijt}. In the asymptotic region, by analyzing the effective potential wall in Hamiltonian, the description of the Kasner transformation can be established by an algebraic method, which is very helpful for the understanding of hidden  symmetries~\cite{Damour:2000hv}. Similar to the analysis of effective potential wall in the Hamiltonian describing Kasner transformation, we have obtained the dynamics by analyzing the non-integrable terms in the equations of motion. Our method not only provides the transformation laws analytically, but also gives the key differential equations that characterize the transformation process. These differential equations are analytically solvable and completely consistent with the results obtained by numerically solving the complete equations of motion, see Fig.~\ref{5DTransition} for a direct comparison of the analytical description and the numerical solutions.

Based on the observation from~\eqref{feedF} and~\eqref{alphaV}, we are able to provide some common features about the interior dynamics under certain conditions. In particular, we recovered the "bounce" interior for hairy black holes of AdS gravity coupled to a neutral scalar with a strong even scalar potential~\cite{Hartnoll:2022rdv}. Moreover, we have shown the significant difference between the even and odd super-exponential scalar potentials, see Fig.~\ref{SuperVKasner}. We also provided one example with a negative even super-exponential potential for which no Kanser structure can develop (see Fig.~\ref{SuperVBreak}).
After the $U(1)$ gauge field is introduced, we have found some novel internal dynamics, including the presence of non-Kasner epochs and the random change of the amplitude of the Kasner exponent at late interior times, see Fig.~\ref{FSinhVExp}.

Our current analysis covers many top-down models from superstring/supergravity, thus allowing one to further explore the process of black holes moving towards the singularity in a controllable way and to understand the holographic significance of internal dynamics from the perspective of dual field theory.
While we have revealed some common features about the interior of dynamics, we believe that further investigation of the parameter space could yield other regions of interest. 
In particular, the charged black hole with a super-exponential scalar potential would show very rich internal behaviors that are far from being understood.

We have paid attention to the static black holes with scalar hair. It will be interesting to generalize our discussion to stationary cases and even dynamic black holes, see a recent study on the internal structure of hairy rotating black holes in three dimensions~\cite{Gao:2023rqc}. Our analysis method could be used to understand the interior dynamics of black holes with other matter content, for example, anisotropic black holes with vector hair~\cite{Cai:2021obq,Sword:2022oyg}. We have been limited to the case with $Z=1$, \emph{i.e.} no direct coupling of $\psi$ to $F_{\mu\nu}F^{\mu\nu}$. In fact, our preliminary analysis suggests that some choice of $Z(\psi)$ could strongly affect the interior dynamics, which lies beyond the scope of current work. It is desirable to understand these features in the future.

\section{Acknowledgements}
This work was supported by National Key Research and Development Program of China Grant No. 2020YFC2201501, and the National Natural Science Foundation of China Grants No.12122513, No.12075298, No.11991052 and No.12047503.

\bibliographystyle{JHEP}

\providecommand{\href}[2]{#2}\begingroup\raggedright
\end{document}